\documentclass[reprint, amsmath,amssymb, aps,superscriptaddress]{revtex4-2}

\usepackage{graphicx}
\usepackage{dcolumn}
\usepackage{bm}
\usepackage{nomencl}
\usepackage{amsmath, amssymb} 
\usepackage{breqn} 
\usepackage{empheq,etoolbox} 
\makenomenclature
\usepackage{xargs}                      
\usepackage[pdftex,dvipsnames]{xcolor}  
\usepackage{hyperref}

\begin{document}

\newcommand{\gabriel}[1]{\textbf{\textcolor{orange}{#1}}}
\newcommand{\paul}[1]{\textbf{\textcolor{red}{#1}}}
\newcommand{\charles}[1]{\textbf{\textcolor{violet}{#1}}}
\newcommand{\lct}{\Lambda_{\rm CT}}
\newcommand{\lcs}{\Lambda_{\rm CS}}
\newcommand{\dimlessnb}{\mathcal{N}_{\rm SE}}

\newcommand{\ladhyx}{Laboratoire d'Hydrodynamique (LadHyX), CNRS, Ecole Polytechnique, Institut Polytechnique de Paris, 91120 Palaiseau, France}
\newcommand{\pasteur}{Institut Pasteur, Université Paris Cité, Physical microfluidics and Bioengineering, 25-28 Rue du Dr Roux, 75015 Paris, France}
\newcommand{\phenix}{Laboratoire PHENIX, CNRS, Sorbonne Université, 4 Place Jussieu, 75005 Paris, France}

\title{Motion and hydrodynamic resistance of an elastic bead confined in a square microchannel}

\author{Charles Paul Moore}
 \affiliation{\ladhyx}
 \affiliation{\pasteur}

\author{Hiba Belkadi}
 \affiliation{\ladhyx}
 \affiliation{\pasteur}

\author{Brouna Safi}
 \affiliation{\ladhyx}

\author{Gabriel Amselem}
\affiliation{\ladhyx}

\author{Charles N. Baroud}
 \email{charles.baroud@polytechnique.edu }
 \affiliation{\ladhyx}
 \affiliation{\pasteur}

\date{\today}
\begin{abstract}
Cells and other soft particles are often forced to flow in confined geometries in both laboratory and natural environments, where the elastic deformation induces an additional drag and pressure drop across the particle. In contrast with other multiphase flows, the physical parameters that determine this additional pressure are still not known. Here we start by measuring the pressure drop across a single spherical hydrogel particle as it flows in a microfluidic comparator. This pressure is found to depend on the amount of confinement, elastic modulus, fluid viscosity and velocity. A model for the force balance on the particle is then proposed, by incorporating the above ingredients and relying on simulations of bead geometry and lubrication flow considerations. The final model collapses the force measurements onto a single scaling law spanning several decades, while providing physical insights by recalling elements from classic multiphase flows and contact mechanics.

\end{abstract}

\maketitle

The motion and resistance of multiphase flows in confined geometries is a classic problem of fluid mechanics that has been studied extensively for the case of an air bubble or a liquid droplet entering into a capillary tube filled with an immiscible liquid~\cite{taylor_deposition_1961}. In this flow the equilibria between viscous stresses and surface tension determine the shape of the deformable interface and the thickness of the wetting layer~\cite{bretherton_motion_1961}. This balance of forces in turn determines the mobility of the drop or bubble and the amount of fluid left behind~\cite{schwartz_motion_1986,wong_motion_1995}. Beyond classical studies of the subject, interest has been rekindled with the emergence of droplet microfluidics, where the motion of drops in square or rectangular capillaries raised questions of practical importance about the formation and transport of drops and bubbles in microchannels~\cite{baroud_dynamics_2010,jakiela2012discontinuous}. In all of the above studies, the balance between viscosity and capillary forces means that the physics is determined by a single parameter, the capillary number ${Ca}=\mu U/\gamma$, where $\mu$ is a typical value of viscosity, $U$ a typical value of velocity, and $\gamma$ is the interfacial tension~\cite{fairbrother35}. 

A different multiphase situation has emerged recently in microfluidic applications but has received much less attention, namely the flow of a soft solid bead in a confined geometry. Such particles can represent the passage of cells in microchannels~\cite{shelby_microfluidic_2003,abkarian_high-speed_2006}, or they can be encountered in situations in which hydrogel beads are co-encapsulated with cells inside droplets~\cite{abate_efficient_2011}. In this second case, it was shown that the droplet production could be synchronized with the microgel bead ejection in order to yield a deterministic encapsulation of a single bead per droplet~\cite{abate_beating_2009}. In contrast with drops and bubbles however, the soft particles do not introduce any surface tension nor do they lead to recirculating patterns inside or outside. Instead, their transport is expected to be dictated by the viscosity of the carrier fluid, the elastic properties of the beads, and the amount of confinement.

Even though the physics of soft particle transport has been studied~\cite{preira_single_2013,khan_passage_2017}, we are still lacking a scaling law that explains how the physical and confinement parameters combine together to determine the bead velocity. This is done below to determine the additional force required to flow the bead with the fluid. 

\begin{figure}[ht!]
    \centering
    \includegraphics[width=\columnwidth]{Figures/Figure_Methods_V0.5.pdf}
    \caption{
    (a) An isometric representation of the microchannel with a bead causing a deflection in the ink tracer. The direction of inlet flow, $Q$ is indicated, as are the test channel width, and height, $w=100~\mu$m  $h=97~\mu$m respectively.
    (b) A micrograph of the microfluidic comparator used in this study, showing the deflection of the ink tracer at the comparator entrance, $x_{\rm cr}$. Flow is passing from left to right.
    (c) Progression of a bead (circled in red) as it enters the comparator and flows into the test channel. Scale bars: $100~\mu$m.}
    \label{fig:Methods}
\end{figure}

The experiments were conducted by flowing previously formed microgel beads into a microfluidic comparator, as shown in Fig.~\ref{fig:Methods}a. It consisted of two identical parallel channels of length $l=2 \; \rm mm$ and width $w= 100 \; \mu \rm m$, and height $h=97 \;\mu \rm m$ that were connected both upstream and downstream of the test section (see Fig.~\ref{fig:sup_comparator}~\cite{moore_supplementary_nodate}). This  comparator geometry has previously been shown to provide  a measurement of the change in pressure between the condition before the bead's arrival and when it is flowing in the channel~\cite{abkarian_high-speed_2006,vanapalli_hydrodynamic_2009}. Flow into the device was controlled using a pair of syringe pumps, one containing a dilute suspension of microgel beads and the other a dyed fluid. The microgels consisted of polyethylene-glycol based gels, with diameters  $d \in ( 80 , 140)~\mu$m and elastic moduli $E^{\star} \in ( 10^3 , 10^5)$~Pa~\cite{moore_clogging_2023}. They were suspended in either a water and surfactant solution or in a glycerol solution, with dynamic viscosities in the range $\mu \in (10^{-3} , 10^{-2})$~Pa.s. The applied flow rates were in the range $Q=5-200~\mu$l/min, such that the Reynolds number $Re={u_{\rm fluid} w \rho}/{\mu} \in (0.04,46)$, with $u_{\rm fluid}$ the average fluid velocity. Details on the fabrication of the microgel beads, measurement of elasticity, and measurement of the suspending fluid viscosity can be found in Ref.~\cite{moore_clogging_2023} and in the {Supplementary Material A,B}~\cite{moore_supplementary_nodate}.

At the beginning of an experiment, the flow rates of both fluids are fixed such that the interface between them lies in the middle of the comparator in the absence of a gel bead. When a bead enters the test channel, it increases the hydrodynamic resistance of the channel, therefore modifying the flow rate distribution between the test and bypass channels by an amount $\Delta Q$. The resulting change in relative flow rates deflects the interface between the two fluids in the comparator region by a value $x_{\rm cr}$, as shown in Fig.~\ref{fig:Methods}b. The calculation of $\Delta Q = f(x_{\rm cr}/w_{\rm cr})$ can be found in the Supplementary material B~\cite{moore_supplementary_nodate}.
The additional pressure drop in the test channel due to the bead can then be calculated as $P_{\rm gel} = 2\Delta Q R_{\rm 0}$, where $R_0=28.5\frac{\mu l}{w^4}$ is the hydrodynamic resistance of identical bypass and test channels in the absence of beads ~\cite{vanapalli_scaling_2007,vanapalli_hydrodynamic_2009} (see Supplementary material B~\cite{moore_supplementary_nodate} for a derivation of the relation between $P_{\rm gel}$, $x_{\rm cr}$ and $\Delta Q$). 
In addition to the ink deflection, the bead velocity $u_{\rm gel}$ is simultaneously measured by tracking the bead position over time, $z_{\rm gel}$, as shown in Fig.~\ref{fig:Methods}c. As the bead velocity reaches a constant value shortly after the bead enters into the test channel, the values of $u_{\rm gel}$ and $x_{\rm cr}$ shown below correspond to the steady-state value observed for each experiment (see Supplementary Material C for full details ~\cite{moore_supplementary_nodate}). 

\begin{figure}[ht!]
    \centering
    \includegraphics{Figures/Figure_PressureVelocityRaw_V0.7.pdf}
    \caption{
    (a) The pressure drop across the bead $P_{\rm gel}$ as a function of the bead confinement, $d/w$. Symbols represent the combinations of microgel bead elasticities $E^{\star}$ and either water or glycerol based carrier fluids. 
    (b) The relative pressure drop of the added bead to that of an empty channel, $\frac{P_{\rm gel}w^2}{28.5 \mu u_{\rm fluid} l}$, as a function of the bead confinement, $d/w$. Symbols represent the combinations of bead elasticities $E^{\star}$ and either water or glycerol based carrier fluids. Symbols and colors correspond to those in part a.
    (c) The relative velocity of beads passing through the test channel, $u_{\rm gel}/u_{\rm fluid}$ for different values of the bead confinement, $d/w$. Colors indicate the pressure drop normalized with respect to bead elasticity, $P_{\rm gel}/E^{\star}$. Inset: the bead velocity $u_{\rm gel}$ vs. mean fluid velocity  $u_{\rm fluid}$. The identity line ($y=x$) is shown for reference.
}
    \label{fig:Results}
\end{figure}

The measurements confirm that the added pressure $P_{\rm gel}$ depends on two physical parameters and on the amount of confinement, as shown in Fig.~\ref{fig:Results}a. Indeed, varying the confinement $d/w$ by a factor $\approx 2$ leads to an increase in pressure of two orders of magnitude. The elasticity of the hydrogel also influences the added pressure $P_{\rm gel}$: increasing the Young modulus from $E^{\star} \approx 10^3$~Pa to $E^{\star} \approx 10^5$~Pa gives rise to an increase of approximately an order of magnitude in $P_{\rm gel}$. Last, for a given gel elasticity, increasing the viscosity of the carrier fluid by an order of magnitude leads to an increase of about an order of magnitude in the pressure needed to push the gel forward.

The effect of confinement on pressure can also be considered from the point of view of the added resistance of the bead itself, as shown in Fig.~\ref{fig:Results}b. Considering that the resistance of an empty microchannel is given by $R_0$, the relative resistance of the bead can be determined from $\frac{\Delta R}{R_0}=\frac{P_{\rm gel} w^2}{28.5 u_{\rm fluid} \mu l}$. Although resistance owing to the bead increases along with degree of confinement, the added resistance remains smaller than that of the microchannel itself, that is, $\Delta R/R_0<1$ in the vast majority of cases.

Measurements of the bead velocity, $u_{\rm gel}$, remain close to $u_{\rm fluid}$, as shown in the inset of Fig.~\ref{fig:Results}c. Close inspection of the beads relative motion however,  $u_{\rm gel}/u_{\rm fluid}$, shows a dependence on the confinement parameter $d/w$, with unconfined beads flowing faster than the fluid velocity and a decreased relative motion for larger confinement. 
Furthermore, the measurements in Fig.~\ref{fig:Results}c reveal that the relative motion of the beads tends to increase with $P_{\rm gel}/E^{\star}$, for a given ratio of $d/w$, as seen in the shading of the color of the points. This dependence is not straightforward to interpret, since $P_{\rm gel}$ also increases with $d/w$. We can therefore only note that the bead velocity is coupled to the bead pressure, while being influenced by the confinement and fluid velocity. The theoretical model developed below provides a quantitative understanding of this coupling.

\begin{figure}
    \centering
    \includegraphics{Figures/Figure_SimResults_V0.11.pdf}
    \caption{
    (a) A micrograph of a flowing bead, with the simulated volume highlighted in red. A schematic showing the confined bead is shown below with two cut planes of symmetry in blue, marking the midpoint between the top and side channel walls. Finally, the simulated volume is shown below this, with the bead highlighted in blue, and the channel walls in grey.
    (b) A diagram of the simulated bead, blue, surrounded by the channel walls, gray, as well as the empty space between, the gutter, highlighted. The cross sectional area, $\lcs$, is represented by the visible area of the simulated bead in blue. 
    (c) An isometric view of the simulated bead in undeformed, left, and deformed, right, is outlined, with color representing contact pressure. The symmetry cut plane of the bead is left in gray. 
    (d) The normalized cross sectional area of a bead, $\lcs/w^2$ vs. the bead confinement, $d/w$. 
    The solid line corresponds to Eq.~\eqref{eq:CrossArea}. 
    (e) The normalized simulated contact area of a bead, $\lct/d^2$ vs. bead confinement, $d/w$. The solid line corresponds to Eq.~\eqref{eq:ContactArea}.
    }
    \label{fig:ContactSim}
\end{figure}

To understand the effect that confinement has on the bead passage, we simulated the geometry of confined beads using the Abaqus CAE software. The simulations consisted in placing an elastic spherical bead of diameter $d$ and Young's modulus $E^{\star}$ in a microchannel with a square cross section of width $w_0$, as shown in Fig.~\ref{fig:ContactSim}a. The walls of the microchannel were assumed to be infinitely rigid. The Poisson ratio of the microgel was chosen to be $\nu=0.35$, a typical value for hydrogels~\cite{chappel_review_2020}, but it was not found that the exact value of $\nu$ played an important role in the following results. The channel dimensions were then reduced by an amount $\Delta w$, while preserving the square cross section, and the deformed shape of the bead was calculated for different values of $d/w$, as shown in Fig.~\ref{fig:ContactSim}b-c. For ease of simulation, problem symmetries were invoked to calculate a 1/4 model of the gel and channel, i.e. cutting the particle along its symmetry axes and only considering this section, as illustrated in Fig.~\ref{fig:ContactSim}a. 

Two surface areas are of particular interest and are obtained from these simulations. First, we measure the cross-sectional area $\lcs$ occupied by the gel within the microchannel, taken in the middle of the bead in the direction of the channel (see Fig.~\ref{fig:ContactSim}b). As confinement increases, $\lcs/w^2$ is also found to increase, as shown in Fig.~\ref{fig:ContactSim}d. This cross-sectional area is geometrically bounded by $\lcs=\frac{\pi}{4}d^2$ in the case of a small unconfined bead, where $d/w = 1$, and $\lcs \rightarrow w^2$ for a highly confined bead, where $d/w = \sqrt{2}$. Then fitting the data with a quadratic relationship between $\lcs$ and $d/w$ that satisfies  the above boundary conditions leads to an analytical approximation
\begin{equation}
    \lcs = \left[-1.25 \left(\frac{d}{w}\right)^2+3.54\left(\frac{d}{w}\right)-1.50\right] w^2.
    \label{eq:CrossArea}
\end{equation}
This solution is found to be in good agreement with the  simulated data in Fig.~\ref{fig:ContactSim}d ($R^2=0.98$).

The second surface of interest concerns the contact surface between the bead and the channel walls, $\lct$, as shown in Fig.~\ref{fig:ContactSim}c. It is known from classical contact mechanics that when an elastic sphere of diameter $d$ is pressed upon a flat surface, the area of contact between the two solids scales as $\Lambda_{\rm Hertz} = {\pi} d p/2$, where $p$ is the penetration of the sphere into the flat surface~\cite{barber_contact_2018}. In the case simulated here, where the bead is squeezed between four walls, $p=(d-w)/2$. 
For contact between two axes, \emph{i.e.} confined on four sides,  the volume of the bead must be pushed outward in the direction of the channel. For an isotropic material with $0 < \nu < 0.5$, the compression in one direction will be proportional to the perpendicular extension, meaning that due to the second axis of confinement, the particle will further elongate in the direction of the channel proportionately to $d/w$. 
Multiplying this proportionality with the classical point contact equation results in the contact area
\begin{equation}
    \lct =  \frac{\pi}{4} \frac{d^2}{w} \left(d-w\right).
    \label{eq:ContactArea}
\end{equation}
This relationship is shown in Fig.~\ref{fig:ContactSim}e. 
The modeled contact area is also in excellent agreement with the simulated data ($R^2=0.99$), and is therefore used here.

The two areas above now allow us to estimate the different forces acting on the bead as it flows. 
In the direction of the flow, we recall that the bead quickly reaches a constant velocity during its passage in the comparator (see  {Supplementary Materials C}~\cite{moore_supplementary_nodate}). This indicates that the sum of forces acting on it in the streamwise direction is zero, meaning that there is a balance between the force pushing the bead forward and the frictional force slowing it down. The force that pushes the bead forward is due to pressure drop across the gel and can be calculated as $F_{\rm P}=P_{\rm gel}\lcs$, where $P_{\rm gel}$ is obtained from the experimental measurements. 

Meanwhile, two different domains on the bead surface experience friction: a gutter domain in the corners of the channel, where the fluid moves past the particle, and a lubricated surface, where the bead is closest to contacting the microchannel. Assuming the transit of the bead only minimally deforms the particle beyond the dry deformation shown in Fig.~\ref{fig:ContactSim}b-c, the lubricated area can be approximated by $\lct$. Friction due to flow in the gutters between the bead and the corner is negligible, since the bead and fluid velocities are similar in all our experiments, the magnitude of $u_{\rm gel}-u_{\rm fluid}$ is small. The shear force on the surface of the bead is therefore small, and so the gutter friction is $|F_{\rm f,\;gutter}| < |0.1\; F_{\rm P}|$ for 95 \% of the data. A quantitative estimate of fluid friction in the gutter region is expanded upon in the {Supplementary Materials} \cite{moore_supplementary_nodate}.

In contrast to gutter friction, a large friction force emerges in the lubrication layers,  where the bead is most nearly in contact with the microchannel wall and entrains the fluid along with it~\cite{hamrock_fundamentals_2004}. The friction force can be calculated as
\begin{equation}
    F_{\rm f}= 4\int_\lct{\sigma (x,z)}dxdz,
    \label{eq:FrictionDef}
\end{equation}
where $\sigma$ is the shear stress acting on $\lct$. We can approximate the shear stress as $\sigma \approx \mu u_{gel}/t$, letting $t$ be the local thickness of the lubricating fluid, as illustrated in Fig.~\ref{fig:EHL_Scaling}a. To estimate $t$, we turn to the equations of elastohydrodynamics.

When an elastic bead flows above a rigid surface, the thickness of the lubrication layer $t(x,z)$ depends on the pressure in the lubricating film $P(x,z)$ and on the Young modulus of the bead~\cite{zargari2007investigation, hamrock_isothermal_1976, hamrock_fundamentals_2004}:
\begin{equation}
    t(x,z)=t_{\rm 0}+2 \frac{\pi}{E^{\star}} \int{\frac{P(x',z')}{\sqrt{(x-x')^2+(z-z')^2}}}dx'dz',
    \label{eq:Hertz}
\end{equation}
where $t_0$ represents the undeformed shape of the bead, as sketched in Fig.~\ref{fig:EHL_Scaling}a. Moreover the pressure $P(x,z)$ and the height $t(x,z)$ of the lubricating film are also related through the Reynolds equation~\cite{guyon2015physical}, which describes the pressure field related to the bead motion:
\begin{equation}
    \frac{\partial}{\partial x}\left(t^3\frac{\partial P}{\partial x} \right)+\frac{\partial}{\partial z}\left( t^3 \frac{\partial P}{ \partial z} \right)=6\mu u_{\rm gel}\frac{\partial t}{\partial x}.
    \label{eq:Reynolds}
\end{equation}

\begin{figure}[ht]
    \centering
    \includegraphics{Figures/Figure_EHLAnalysis_V0.12.pdf}
    \caption{(a) A diagram showing the bead (cyan) nearly in contact with the microchannel walls. The distance from the bead to the bead wall, $t$ is highlighted in the blowup, as is the lubricating point pressure, $P$.
    (b)     A comparison of the pressure driving the bead with the predicted frictional scaling due to the lubricating layer. The pressure is normalized by a term expected to scale with the shear acting on the bead surface.
    Data are segregated by elasticity and viscosity. A unity line is provided for reference. Symbols and colors correspond to those in Fig.~\ref{fig:Results}a.
    }
    \label{fig:EHL_Scaling}
\end{figure}

The resulting Eqs.~\eqref{eq:Hertz} and~\eqref{eq:Reynolds} constitute a coupled set of differential equations for $P$ and $t$ that can only be solved numerically~\cite{hamrock_fundamentals_2004}. However their treatment can be greatly simplified by considering the scaling behavior and then injecting the scaling relation for $P$ and $t$ into Eq.~\eqref{eq:FrictionDef}. This yields an estimate of the friction force as a function of the physical and geometric parameters:
\begin{equation}
    F_{\rm f} \sim  \left( \mu u_{\rm gel} \lct \right)^{2/3} E^{* 1/3}.
    \label{eq:fricscaling}
\end{equation}
The full derivation of this scaling law can be found in the {Supplementary Materials}\cite{moore_supplementary_nodate}.

Finally, writing a force balance between the pressure and friction forces $F_{\rm P}=F_{\rm f}$ provides a scaling of the pressure due to the presence of the gel bead, as a function of its velocity and the physical and geometric parameters:
\begin{equation}
\begin{aligned}
       \frac{P_{\rm gel}w}{u_{\rm gel}\mu} \sim \left(\frac{E^* w}{u_{\rm gel}\mu}\right)^{1/3} \frac{(\lct w)^{2/3}}{\lcs}.
    \label{eq:FfVsFp}
\end{aligned}
\end{equation}

Here, the added pressure due to the presence of the bead is normalized by the representative viscous shear rate for a mean velocity $u_{gel}$. Rescaling the data according to Eq.~\eqref{eq:FfVsFp} yields a very good collapse, as shown in Fig.~\ref{fig:EHL_Scaling}b. The slope of the collapsed data agrees well with the theoretical scaling over more than one decade. This collapse is notable since the experiments cover a wide range of values of fluid and solid properties, as well as a wide range of Reynolds numbers and pressure values. 

Inertial effects may be expected to play a role at high values of the Reynolds number, for instance if the bead passage time becomes comparable with the time for inertial effects to decay. From the data we find that the passage time through the comparator, for the fastest flowing beads, is only about 4~ms. This value is comparable with the viscous diffusion time across the half width of the channel, which is $\tau_{\rm visc}=(w/2)^2\rho/\mu\simeq 2.5$~ms. While a modified model to account for transient effects would require much more complex analysis, such inertial effects may explain the departure of the left-most triangles in Fig.~\ref{fig:EHL_Scaling}b from the general scaling law. Other physical effects may also influence the data collapse, including poro-elasticity effects in the moving beads, or the deformability of the microchannels at high driving pressures~\cite{christov_soft_2022,guyard_elastohydrodynamic_2022,tavakol_extended_2017}. However the current protocols do not allow us to address these higher-order phenomena.

It is instructive to compare this scaling law with more sophisticated models of soft lubrication and elasto-hydrodynamic lubrication~\cite{hamrock_isothermal_1976,rallabandi_fluid-elastic_2024}. This subject has been extensively studied in the tribology literature, particularly for a spherical or cylindrical bead being moved along a flat planar surface, with one or both materials being soft~\cite{cartas_ayala_hydrodynamic_2013, sadowski_friction_2019,zhao_modelling_2019, moghani_determinants_2009, de_vicente_frictional_2005, zakhari_slip_2021, peng_elastohydrodynamic_2021, meeker_slip_2004,bongaerts_soft-tribology_2007}. The elastic-viscous balance is used in many of those models to determine the shape of the lubrication layer~\cite{snoeijer_similarity_2013} and its relation with the frictional force~\cite{rallabandi_fluid-elastic_2024}, similarly to what we present here. The frictional force that is calculated for these unconfined studies is frequently found to scale with a power law of the physical parameters as $F_f\sim \left(\mu u\right)^n E^{*(1-n)}$, with $n$ ranging from 0.3~\cite{de_vicente_soft_2006} to 0.65~\cite{cartas_ayala_hydrodynamic_2013}. Our scaling ($n=2/3$) lies in the upper bound of this range, but is consistent with existing findings.

A major difference however between the unconfined case and our microchannel measurements is that the bead velocity in the unconfined cases is imposed by the experimental apparatus, rather than emerging from an equilibrium between a carrier fluid flow and wall friction. As such the three-way adaptation between the bead velocity, its shape, and the friction is more reminiscent of the flow of drops or bubbles in square channels~\cite{baroud_dynamics_2010}, where the pressure drop varies with the capillary number as $\Delta P \sim {\rm Ca}^{2/3}$~\cite{wong_motion_1995,wong_motion_1995-1,baudoin_airway_2013}. Although the physical ingredients are different between drops and elastic beads, both the transit of bubbles and elastic bodies involve balancing a driving pressure that acts on the scale of the channel ($F_p$ here) with a drag force that is dominated by velocity gradients within a lubrication layer ($F_f$ here). The scaling law that is emerges from this force balance (Eq.~\ref{eq:FfVsFp}) combines a measure of the confinement with a nondimensional term that accounts for this physical equilibrium in a way that recalls the Capillary number for droplet or bubble flow in channels.

The new physical understanding obtained from the current experiments and theory can be applied in a variety of microfluidic applications. For example, the transport of cells in confined microfluidic channels has been used to measure their mechanical properties, although most of the modeling has been based on complex numerical work~\cite{shelby_microfluidic_2003,preira_single_2013, luo_constriction_2014, xue_constriction_2015}. The scaling shown here can be used to relate the physical properties without the need for complex numerics. From a technological point of view, the controlled encapsulation of individual soft particles in microfluidic droplets has become a key enabler aspect in single cell genetic analysis ~\cite{zilionis_single-cell_2016}. Although the presence of beads downstream of the droplet production region has been evoked as the source of deterministic encapsulation~\cite{abate_beating_2009}, the results shown here provide a quantitative physical insight into the limits and opportunities of such downstream interactions.

\bibliographystyle{apsrev4-2}
\bibliography{Moore_Bibliography_Export.bib,moreBiblio.bib,ExtraRef_FlowingBeads.bib}

\begin{thebibliography}{50}%
\makeatletter
\providecommand \@ifxundefined [1]{%
 \@ifx{#1\undefined}
}%
\providecommand \@ifnum [1]{%
 \ifnum #1\expandafter \@firstoftwo
 \else \expandafter \@secondoftwo
 \fi
}%
\providecommand \@ifx [1]{%
 \ifx #1\expandafter \@firstoftwo
 \else \expandafter \@secondoftwo
 \fi
}%
\providecommand \natexlab [1]{#1}%
\providecommand \enquote  [1]{``#1''}%
\providecommand \bibnamefont  [1]{#1}%
\providecommand \bibfnamefont [1]{#1}%
\providecommand \citenamefont [1]{#1}%
\providecommand \href@noop [0]{\@secondoftwo}%
\providecommand \href [0]{\begingroup \@sanitize@url \@href}%
\providecommand \@href[1]{\@@startlink{#1}\@@href}%
\providecommand \@@href[1]{\endgroup#1\@@endlink}%
\providecommand \@sanitize@url [0]{\catcode `\\12\catcode `\$12\catcode `\&12\catcode `\#12\catcode `\^12\catcode `\_12\catcode `\%12\relax}%
\providecommand \@@startlink[1]{}%
\providecommand \@@endlink[0]{}%
\providecommand \url  [0]{\begingroup\@sanitize@url \@url }%
\providecommand \@url [1]{\endgroup\@href {#1}{\urlprefix }}%
\providecommand \urlprefix  [0]{URL }%
\providecommand \Eprint [0]{\href }%
\providecommand \doibase [0]{https://doi.org/}%
\providecommand \selectlanguage [0]{\@gobble}%
\providecommand \bibinfo  [0]{\@secondoftwo}%
\providecommand \bibfield  [0]{\@secondoftwo}%
\providecommand \translation [1]{[#1]}%
\providecommand \BibitemOpen [0]{}%
\providecommand \bibitemStop [0]{}%
\providecommand \bibitemNoStop [0]{.\EOS\space}%
\providecommand \EOS [0]{\spacefactor3000\relax}%
\providecommand \BibitemShut  [1]{\csname bibitem#1\endcsname}%
\let\auto@bib@innerbib\@empty
\bibitem [{\citenamefont {Taylor}(1961)}]{taylor_deposition_1961}%
  \BibitemOpen
  \bibfield  {author} {\bibinfo {author} {\bibfnamefont {G.~I.}\ \bibnamefont {Taylor}},\ }\href {https://doi.org/10.1017/S0022112061000159} {\bibfield  {journal} {\bibinfo  {journal} {Journal of Fluid Mechanics}\ }\textbf {\bibinfo {volume} {10}},\ \bibinfo {pages} {161} (\bibinfo {year} {1961})}\BibitemShut {NoStop}%
\bibitem [{\citenamefont {Bretherton}(1961)}]{bretherton_motion_1961}%
  \BibitemOpen
  \bibfield  {author} {\bibinfo {author} {\bibfnamefont {F.~P.}\ \bibnamefont {Bretherton}},\ }\href {https://doi.org/10.1017/S0022112061000160} {\bibfield  {journal} {\bibinfo  {journal} {Journal of Fluid Mechanics}\ }\textbf {\bibinfo {volume} {10}},\ \bibinfo {pages} {166} (\bibinfo {year} {1961})}\BibitemShut {NoStop}%
\bibitem [{\citenamefont {Schwartz}\ \emph {et~al.}(1986)\citenamefont {Schwartz}, \citenamefont {Princen},\ and\ \citenamefont {Kiss}}]{schwartz_motion_1986}%
  \BibitemOpen
  \bibfield  {author} {\bibinfo {author} {\bibfnamefont {L.~W.}\ \bibnamefont {Schwartz}}, \bibinfo {author} {\bibfnamefont {H.~M.}\ \bibnamefont {Princen}},\ and\ \bibinfo {author} {\bibfnamefont {A.~D.}\ \bibnamefont {Kiss}},\ }\href {https://doi.org/10.1017/S0022112086001738} {\bibfield  {journal} {\bibinfo  {journal} {Journal of Fluid Mechanics}\ }\textbf {\bibinfo {volume} {172}},\ \bibinfo {pages} {259} (\bibinfo {year} {1986})}\BibitemShut {NoStop}%
\bibitem [{\citenamefont {Wong}\ \emph {et~al.}(1995{\natexlab{a}})\citenamefont {Wong}, \citenamefont {Radke},\ and\ \citenamefont {Morris}}]{wong_motion_1995}%
  \BibitemOpen
  \bibfield  {author} {\bibinfo {author} {\bibfnamefont {H.}~\bibnamefont {Wong}}, \bibinfo {author} {\bibfnamefont {C.~J.}\ \bibnamefont {Radke}},\ and\ \bibinfo {author} {\bibfnamefont {S.}~\bibnamefont {Morris}},\ }\href {https://doi.org/10.1017/S0022112095001443} {\bibfield  {journal} {\bibinfo  {journal} {Journal of Fluid Mechanics}\ }\textbf {\bibinfo {volume} {292}},\ \bibinfo {pages} {71} (\bibinfo {year} {1995}{\natexlab{a}})}\BibitemShut {NoStop}%
\bibitem [{\citenamefont {Baroud}\ \emph {et~al.}(2010)\citenamefont {Baroud}, \citenamefont {Gallaire},\ and\ \citenamefont {Dangla}}]{baroud_dynamics_2010}%
  \BibitemOpen
  \bibfield  {author} {\bibinfo {author} {\bibfnamefont {C.~N.}\ \bibnamefont {Baroud}}, \bibinfo {author} {\bibfnamefont {F.}~\bibnamefont {Gallaire}},\ and\ \bibinfo {author} {\bibfnamefont {R.}~\bibnamefont {Dangla}},\ }\href {https://doi.org/10.1039/c001191f} {\bibfield  {journal} {\bibinfo  {journal} {Lab on a chip}\ }\textbf {\bibinfo {volume} {1}},\ \bibinfo {pages} {232} (\bibinfo {year} {2010})}\BibitemShut {NoStop}%
\bibitem [{\citenamefont {Jakiela}\ \emph {et~al.}(2012)\citenamefont {Jakiela}, \citenamefont {Korczyk}, \citenamefont {Makulska}, \citenamefont {Cybulski},\ and\ \citenamefont {Garstecki}}]{jakiela2012discontinuous}%
  \BibitemOpen
  \bibfield  {author} {\bibinfo {author} {\bibfnamefont {S.}~\bibnamefont {Jakiela}}, \bibinfo {author} {\bibfnamefont {P.~M.}\ \bibnamefont {Korczyk}}, \bibinfo {author} {\bibfnamefont {S.}~\bibnamefont {Makulska}}, \bibinfo {author} {\bibfnamefont {O.}~\bibnamefont {Cybulski}},\ and\ \bibinfo {author} {\bibfnamefont {P.}~\bibnamefont {Garstecki}},\ }\href@noop {} {\bibfield  {journal} {\bibinfo  {journal} {Physical review letters}\ }\textbf {\bibinfo {volume} {108}},\ \bibinfo {pages} {134501} (\bibinfo {year} {2012})}\BibitemShut {NoStop}%
\bibitem [{\citenamefont {Fairbrother}\ and\ \citenamefont {Stubbs}(1935)}]{fairbrother35}%
  \BibitemOpen
  \bibfield  {author} {\bibinfo {author} {\bibfnamefont {F.}~\bibnamefont {Fairbrother}}\ and\ \bibinfo {author} {\bibfnamefont {A.~E.}\ \bibnamefont {Stubbs}},\ }\href@noop {} {\bibfield  {journal} {\bibinfo  {journal} {Journal of Chemical Society}\ }\textbf {\bibinfo {volume} {1}} (\bibinfo {year} {1935})}\BibitemShut {NoStop}%
\bibitem [{\citenamefont {Shelby}\ \emph {et~al.}(2003)\citenamefont {Shelby}, \citenamefont {White}, \citenamefont {Ganesan}, \citenamefont {Rathod},\ and\ \citenamefont {Chiu}}]{shelby_microfluidic_2003}%
  \BibitemOpen
  \bibfield  {author} {\bibinfo {author} {\bibfnamefont {J.~P.}\ \bibnamefont {Shelby}}, \bibinfo {author} {\bibfnamefont {J.}~\bibnamefont {White}}, \bibinfo {author} {\bibfnamefont {K.}~\bibnamefont {Ganesan}}, \bibinfo {author} {\bibfnamefont {P.~K.}\ \bibnamefont {Rathod}},\ and\ \bibinfo {author} {\bibfnamefont {D.~T.}\ \bibnamefont {Chiu}},\ }\href {https://doi.org/10.1073/pnas.2433968100} {\bibfield  {journal} {\bibinfo  {journal} {Proceedings of the National Academy of Sciences}\ }\textbf {\bibinfo {volume} {100}},\ \bibinfo {pages} {14618} (\bibinfo {year} {2003})}\BibitemShut {NoStop}%
\bibitem [{\citenamefont {Abkarian}\ \emph {et~al.}(2006)\citenamefont {Abkarian}, \citenamefont {Faivre},\ and\ \citenamefont {Stone}}]{abkarian_high-speed_2006}%
  \BibitemOpen
  \bibfield  {author} {\bibinfo {author} {\bibfnamefont {M.}~\bibnamefont {Abkarian}}, \bibinfo {author} {\bibfnamefont {M.}~\bibnamefont {Faivre}},\ and\ \bibinfo {author} {\bibfnamefont {H.~A.}\ \bibnamefont {Stone}},\ }\href {https://doi.org/10.1073/pnas.0507171102} {\bibfield  {journal} {\bibinfo  {journal} {Proceedings of the National Academy of Sciences}\ }\textbf {\bibinfo {volume} {103}},\ \bibinfo {pages} {538} (\bibinfo {year} {2006})}\BibitemShut {NoStop}%
\bibitem [{\citenamefont {Abate}\ \emph {et~al.}(2011)\citenamefont {Abate}, \citenamefont {Rotem}, \citenamefont {Thiele},\ and\ \citenamefont {Weitz}}]{abate_efficient_2011}%
  \BibitemOpen
  \bibfield  {author} {\bibinfo {author} {\bibfnamefont {A.~R.}\ \bibnamefont {Abate}}, \bibinfo {author} {\bibfnamefont {A.}~\bibnamefont {Rotem}}, \bibinfo {author} {\bibfnamefont {J.}~\bibnamefont {Thiele}},\ and\ \bibinfo {author} {\bibfnamefont {D.~A.}\ \bibnamefont {Weitz}},\ }\href {https://doi.org/10.1103/PhysRevE.84.031502} {\bibfield  {journal} {\bibinfo  {journal} {Physical review. E, Statistical, nonlinear, and soft matter physics}\ }\textbf {\bibinfo {volume} {84}},\ \bibinfo {pages} {031502} (\bibinfo {year} {2011})}\BibitemShut {NoStop}%
\bibitem [{\citenamefont {Abate}\ \emph {et~al.}(2009)\citenamefont {Abate}, \citenamefont {Chen}, \citenamefont {Agresti},\ and\ \citenamefont {Weitz}}]{abate_beating_2009}%
  \BibitemOpen
  \bibfield  {author} {\bibinfo {author} {\bibfnamefont {A.~R.}\ \bibnamefont {Abate}}, \bibinfo {author} {\bibfnamefont {C.-H.}\ \bibnamefont {Chen}}, \bibinfo {author} {\bibfnamefont {J.~J.}\ \bibnamefont {Agresti}},\ and\ \bibinfo {author} {\bibfnamefont {D.~A.}\ \bibnamefont {Weitz}},\ }\href {https://doi.org/10.1039/b909386a} {\bibfield  {journal} {\bibinfo  {journal} {Lab on a chip}\ }\textbf {\bibinfo {volume} {9}},\ \bibinfo {pages} {2628} (\bibinfo {year} {2009})}\BibitemShut {NoStop}%
\bibitem [{\citenamefont {Preira}\ \emph {et~al.}(2013)\citenamefont {Preira}, \citenamefont {Valignat}, \citenamefont {Bico},\ and\ \citenamefont {Théodoly}}]{preira_single_2013}%
  \BibitemOpen
  \bibfield  {author} {\bibinfo {author} {\bibfnamefont {P.}~\bibnamefont {Preira}}, \bibinfo {author} {\bibfnamefont {M.-P.}\ \bibnamefont {Valignat}}, \bibinfo {author} {\bibfnamefont {J.}~\bibnamefont {Bico}},\ and\ \bibinfo {author} {\bibfnamefont {O.}~\bibnamefont {Théodoly}},\ }\href {https://doi.org/10.1063/1.4802272} {\bibfield  {journal} {\bibinfo  {journal} {Biomicrofluidics}\ }\textbf {\bibinfo {volume} {7}},\ \bibinfo {pages} {024111} (\bibinfo {year} {2013})}\BibitemShut {NoStop}%
\bibitem [{\citenamefont {Khan}\ \emph {et~al.}(2017)\citenamefont {Khan}, \citenamefont {Kamyabi}, \citenamefont {Hussain},\ and\ \citenamefont {Vanapalli}}]{khan_passage_2017}%
  \BibitemOpen
  \bibfield  {author} {\bibinfo {author} {\bibfnamefont {Z.~S.}\ \bibnamefont {Khan}}, \bibinfo {author} {\bibfnamefont {N.}~\bibnamefont {Kamyabi}}, \bibinfo {author} {\bibfnamefont {F.}~\bibnamefont {Hussain}},\ and\ \bibinfo {author} {\bibfnamefont {S.~A.}\ \bibnamefont {Vanapalli}},\ }\href {https://doi.org/10.1088/2057-1739/aa5f60} {\bibfield  {journal} {\bibinfo  {journal} {Convergent Science Physical Oncology}\ }\textbf {\bibinfo {volume} {3}},\ \bibinfo {pages} {024001} (\bibinfo {year} {2017})}\BibitemShut {NoStop}%
\bibitem [{\citenamefont {Moore}\ \emph {et~al.}(2025)\citenamefont {Moore}, \citenamefont {Belkadi}, \citenamefont {Safi}, \citenamefont {Amselem},\ and\ \citenamefont {Baroud}}]{moore_supplementary_nodate}%
  \BibitemOpen
  \bibfield  {author} {\bibinfo {author} {\bibfnamefont {C.~P.}\ \bibnamefont {Moore}}, \bibinfo {author} {\bibfnamefont {H.}~\bibnamefont {Belkadi}}, \bibinfo {author} {\bibfnamefont {B.}~\bibnamefont {Safi}}, \bibinfo {author} {\bibfnamefont {G.}~\bibnamefont {Amselem}},\ and\ \bibinfo {author} {\bibfnamefont {C.~N.}\ \bibnamefont {Baroud}},\ }\href@noop {} {\bibfield  {journal} {\bibinfo  {journal} {Supplementary Material}\ } (\bibinfo {year} {2025})}\BibitemShut {NoStop}%
\bibitem [{\citenamefont {Vanapalli}\ \emph {et~al.}(2009)\citenamefont {Vanapalli}, \citenamefont {Banpurkar}, \citenamefont {van~den Ende}, \citenamefont {Duits},\ and\ \citenamefont {Mugele}}]{vanapalli_hydrodynamic_2009}%
  \BibitemOpen
  \bibfield  {author} {\bibinfo {author} {\bibfnamefont {S.~A.}\ \bibnamefont {Vanapalli}}, \bibinfo {author} {\bibfnamefont {A.~G.}\ \bibnamefont {Banpurkar}}, \bibinfo {author} {\bibfnamefont {D.}~\bibnamefont {van~den Ende}}, \bibinfo {author} {\bibfnamefont {M.~H.~G.}\ \bibnamefont {Duits}},\ and\ \bibinfo {author} {\bibfnamefont {F.}~\bibnamefont {Mugele}},\ }\href {https://doi.org/10.1039/B815002H} {\bibfield  {journal} {\bibinfo  {journal} {Lab Chip}\ }\textbf {\bibinfo {volume} {9}},\ \bibinfo {pages} {982} (\bibinfo {year} {2009})}\BibitemShut {NoStop}%
\bibitem [{\citenamefont {Moore}\ \emph {et~al.}(2023)\citenamefont {Moore}, \citenamefont {Husson}, \citenamefont {Boudaoud}, \citenamefont {Amselem},\ and\ \citenamefont {Baroud}}]{moore_clogging_2023}%
  \BibitemOpen
  \bibfield  {author} {\bibinfo {author} {\bibfnamefont {C.~P.}\ \bibnamefont {Moore}}, \bibinfo {author} {\bibfnamefont {J.}~\bibnamefont {Husson}}, \bibinfo {author} {\bibfnamefont {A.}~\bibnamefont {Boudaoud}}, \bibinfo {author} {\bibfnamefont {G.}~\bibnamefont {Amselem}},\ and\ \bibinfo {author} {\bibfnamefont {C.~N.}\ \bibnamefont {Baroud}},\ }\href {https://doi.org/10.1103/PhysRevLett.130.064001} {\bibfield  {journal} {\bibinfo  {journal} {Physical Review Letters}\ }\textbf {\bibinfo {volume} {130}},\ \bibinfo {pages} {064001} (\bibinfo {year} {2023})}\BibitemShut {NoStop}%
\bibitem [{\citenamefont {Vanapalli}\ \emph {et~al.}(2007)\citenamefont {Vanapalli}, \citenamefont {van~den Ende}, \citenamefont {Duits},\ and\ \citenamefont {Mugele}}]{vanapalli_scaling_2007}%
  \BibitemOpen
  \bibfield  {author} {\bibinfo {author} {\bibfnamefont {S.~A.}\ \bibnamefont {Vanapalli}}, \bibinfo {author} {\bibfnamefont {D.}~\bibnamefont {van~den Ende}}, \bibinfo {author} {\bibfnamefont {M.~H.~G.}\ \bibnamefont {Duits}},\ and\ \bibinfo {author} {\bibfnamefont {F.}~\bibnamefont {Mugele}},\ }\href {https://doi.org/10.1063/1.2713800} {\bibfield  {journal} {\bibinfo  {journal} {Applied Physics Letters}\ }\textbf {\bibinfo {volume} {90}},\ \bibinfo {pages} {114109} (\bibinfo {year} {2007})}\BibitemShut {NoStop}%
\bibitem [{\citenamefont {Chappel}(2020)}]{chappel_review_2020}%
  \BibitemOpen
  \bibfield  {author} {\bibinfo {author} {\bibfnamefont {E.}~\bibnamefont {Chappel}},\ }\href {https://doi.org/10.3390/app10248858} {\bibfield  {journal} {\bibinfo  {journal} {Applied Sciences}\ }\textbf {\bibinfo {volume} {10}},\ \bibinfo {pages} {8858} (\bibinfo {year} {2020})}\BibitemShut {NoStop}%
\bibitem [{\citenamefont {Barber}(2018)}]{barber_contact_2018}%
  \BibitemOpen
  \bibfield  {author} {\bibinfo {author} {\bibfnamefont {J.}~\bibnamefont {Barber}},\ }\href {https://doi.org/10.1007/978-3-319-70939-0} {\emph {\bibinfo {title} {Contact {Mechanics}}}},\ \bibinfo {series} {Solid {Mechanics} and {Its} {Applications}}, Vol.\ \bibinfo {volume} {250}\ (\bibinfo  {publisher} {Springer International Publishing},\ \bibinfo {address} {Cham},\ \bibinfo {year} {2018})\BibitemShut {NoStop}%
\bibitem [{\citenamefont {Hamrock}\ \emph {et~al.}(2004)\citenamefont {Hamrock}, \citenamefont {Schmid},\ and\ \citenamefont {Jacobson}}]{hamrock_fundamentals_2004}%
  \BibitemOpen
  \bibfield  {author} {\bibinfo {author} {\bibfnamefont {B.~J.}\ \bibnamefont {Hamrock}}, \bibinfo {author} {\bibfnamefont {S.~R.}\ \bibnamefont {Schmid}},\ and\ \bibinfo {author} {\bibfnamefont {B.~O.}\ \bibnamefont {Jacobson}},\ }\href@noop {} {\emph {\bibinfo {title} {Fundamentals of fluid film lubrication}}},\ \bibinfo {edition} {2nd}\ ed.,\ \bibinfo {series} {Mechanical engineering}\ No.\ \bibinfo {number} {169}\ (\bibinfo  {publisher} {Marcel Dekker},\ \bibinfo {address} {New York},\ \bibinfo {year} {2004})\BibitemShut {NoStop}%
\bibitem [{\citenamefont {Zargari}\ \emph {et~al.}(2007)\citenamefont {Zargari}, \citenamefont {Jimack},\ and\ \citenamefont {Walkley}}]{zargari2007investigation}%
  \BibitemOpen
  \bibfield  {author} {\bibinfo {author} {\bibfnamefont {E.~A.}\ \bibnamefont {Zargari}}, \bibinfo {author} {\bibfnamefont {P.}~\bibnamefont {Jimack}},\ and\ \bibinfo {author} {\bibfnamefont {M.}~\bibnamefont {Walkley}},\ }\href@noop {} {\bibfield  {journal} {\bibinfo  {journal} {International Journal for Numerical Methods in Fluids}\ ,\ \bibinfo {pages} {1}} (\bibinfo {year} {2007})}\BibitemShut {NoStop}%
\bibitem [{\citenamefont {Hamrock}\ and\ \citenamefont {Dowson}(1976)}]{hamrock_isothermal_1976}%
  \BibitemOpen
  \bibfield  {author} {\bibinfo {author} {\bibfnamefont {B.~J.}\ \bibnamefont {Hamrock}}\ and\ \bibinfo {author} {\bibfnamefont {D.}~\bibnamefont {Dowson}},\ }\href {https://doi.org/https://doi.org/10.1115/1.3452861} {\bibfield  {journal} {\bibinfo  {journal} {ASME Journal of Lubrication Technology}\ }\textbf {\bibinfo {volume} {98}},\ \bibinfo {pages} {375} (\bibinfo {year} {1976})}\BibitemShut {NoStop}%
\bibitem [{\citenamefont {Guyon}\ \emph {et~al.}(2015)\citenamefont {Guyon}, \citenamefont {Hulin}, \citenamefont {Petit},\ and\ \citenamefont {Mitescu}}]{guyon2015physical}%
  \BibitemOpen
  \bibfield  {author} {\bibinfo {author} {\bibfnamefont {E.}~\bibnamefont {Guyon}}, \bibinfo {author} {\bibfnamefont {J.~P.}\ \bibnamefont {Hulin}}, \bibinfo {author} {\bibfnamefont {L.}~\bibnamefont {Petit}},\ and\ \bibinfo {author} {\bibfnamefont {C.~D.}\ \bibnamefont {Mitescu}},\ }\href@noop {} {\emph {\bibinfo {title} {Physical hydrodynamics}}}\ (\bibinfo  {publisher} {Oxford university press},\ \bibinfo {year} {2015})\BibitemShut {NoStop}%
\bibitem [{\citenamefont {Christov}(2022)}]{christov_soft_2022}%
  \BibitemOpen
  \bibfield  {author} {\bibinfo {author} {\bibfnamefont {I.~C.}\ \bibnamefont {Christov}},\ }\href {https://doi.org/10.1088/1361-648X/ac327d} {\bibfield  {journal} {\bibinfo  {journal} {Journal of Physics: Condensed Matter}\ }\textbf {\bibinfo {volume} {34}},\ \bibinfo {pages} {063001} (\bibinfo {year} {2022})}\BibitemShut {NoStop}%
\bibitem [{\citenamefont {Guyard}\ \emph {et~al.}(2022)\citenamefont {Guyard}, \citenamefont {Restagno},\ and\ \citenamefont {McGraw}}]{guyard_elastohydrodynamic_2022}%
  \BibitemOpen
  \bibfield  {author} {\bibinfo {author} {\bibfnamefont {G.}~\bibnamefont {Guyard}}, \bibinfo {author} {\bibfnamefont {F.}~\bibnamefont {Restagno}},\ and\ \bibinfo {author} {\bibfnamefont {J.~D.}\ \bibnamefont {McGraw}},\ }\href {https://doi.org/10.1103/PhysRevLett.129.204501} {\bibfield  {journal} {\bibinfo  {journal} {Physical Review Letters}\ }\textbf {\bibinfo {volume} {129}},\ \bibinfo {pages} {204501} (\bibinfo {year} {2022})}\BibitemShut {NoStop}%
\bibitem [{\citenamefont {Tavakol}\ \emph {et~al.}(2017)\citenamefont {Tavakol}, \citenamefont {Froehlicher}, \citenamefont {Holmes},\ and\ \citenamefont {Stone}}]{tavakol_extended_2017}%
  \BibitemOpen
  \bibfield  {author} {\bibinfo {author} {\bibfnamefont {B.}~\bibnamefont {Tavakol}}, \bibinfo {author} {\bibfnamefont {G.}~\bibnamefont {Froehlicher}}, \bibinfo {author} {\bibfnamefont {D.~P.}\ \bibnamefont {Holmes}},\ and\ \bibinfo {author} {\bibfnamefont {H.~A.}\ \bibnamefont {Stone}},\ }\href {https://doi.org/10.1098/rspa.2017.0234} {\bibfield  {journal} {\bibinfo  {journal} {Proceedings of the Royal Society A: Mathematical, Physical and Engineering Sciences}\ }\textbf {\bibinfo {volume} {473}},\ \bibinfo {pages} {20170234} (\bibinfo {year} {2017})}\BibitemShut {NoStop}%
\bibitem [{\citenamefont {Rallabandi}(2024)}]{rallabandi_fluid-elastic_2024}%
  \BibitemOpen
  \bibfield  {author} {\bibinfo {author} {\bibfnamefont {B.}~\bibnamefont {Rallabandi}},\ }\href {https://doi.org/10.1146/annurev-fluid-120720-024426} {\bibfield  {journal} {\bibinfo  {journal} {Annual Review of Fluid Mechanics}\ }\textbf {\bibinfo {volume} {56}},\ \bibinfo {pages} {491} (\bibinfo {year} {2024})}\BibitemShut {NoStop}%
\bibitem [{\citenamefont {Cartas~Ayala}(2013)}]{cartas_ayala_hydrodynamic_2013}%
  \BibitemOpen
  \bibfield  {author} {\bibinfo {author} {\bibfnamefont {M.~A.}\ \bibnamefont {Cartas~Ayala}},\ }\emph {\bibinfo {title} {Hydrodynamic resistance and sorting of deformable particles in microfluidic circuits}},\ \href {http://hdl.handle.net/1721.1/79312} {\bibinfo {type} {{PhD} {Thesis}}},\ \bibinfo  {school} {Massachusetts Institute of Technology} (\bibinfo {year} {2013})\BibitemShut {NoStop}%
\bibitem [{\citenamefont {Sadowski}\ and\ \citenamefont {Stupkiewicz}(2019)}]{sadowski_friction_2019}%
  \BibitemOpen
  \bibfield  {author} {\bibinfo {author} {\bibfnamefont {P.}~\bibnamefont {Sadowski}}\ and\ \bibinfo {author} {\bibfnamefont {S.}~\bibnamefont {Stupkiewicz}},\ }\href {https://doi.org/10.1016/j.triboint.2018.08.025} {\bibfield  {journal} {\bibinfo  {journal} {Tribology International}\ }\textbf {\bibinfo {volume} {129}},\ \bibinfo {pages} {246} (\bibinfo {year} {2019})}\BibitemShut {NoStop}%
\bibitem [{\citenamefont {Zhao}\ \emph {et~al.}(2019)\citenamefont {Zhao}, \citenamefont {Zhang},\ and\ \citenamefont {Zhang}}]{zhao_modelling_2019}%
  \BibitemOpen
  \bibfield  {author} {\bibinfo {author} {\bibfnamefont {B.}~\bibnamefont {Zhao}}, \bibinfo {author} {\bibfnamefont {B.}~\bibnamefont {Zhang}},\ and\ \bibinfo {author} {\bibfnamefont {K.}~\bibnamefont {Zhang}},\ }\href {https://doi.org/10.1016/j.triboint.2018.08.042} {\bibfield  {journal} {\bibinfo  {journal} {Tribology International}\ }\textbf {\bibinfo {volume} {129}},\ \bibinfo {pages} {377} (\bibinfo {year} {2019})}\BibitemShut {NoStop}%
\bibitem [{\citenamefont {Moghani}\ \emph {et~al.}(2009)\citenamefont {Moghani}, \citenamefont {Butler},\ and\ \citenamefont {Loring}}]{moghani_determinants_2009}%
  \BibitemOpen
  \bibfield  {author} {\bibinfo {author} {\bibfnamefont {T.}~\bibnamefont {Moghani}}, \bibinfo {author} {\bibfnamefont {J.~P.}\ \bibnamefont {Butler}},\ and\ \bibinfo {author} {\bibfnamefont {S.~H.}\ \bibnamefont {Loring}},\ }\href {https://doi.org/10.1016/j.jbiomech.2009.02.021} {\bibfield  {journal} {\bibinfo  {journal} {Journal of Biomechanics}\ }\textbf {\bibinfo {volume} {42}},\ \bibinfo {pages} {1069} (\bibinfo {year} {2009})}\BibitemShut {NoStop}%
\bibitem [{\citenamefont {de~Vicente}\ \emph {et~al.}(2005)\citenamefont {de~Vicente}, \citenamefont {Stokes},\ and\ \citenamefont {Spikes}}]{de_vicente_frictional_2005}%
  \BibitemOpen
  \bibfield  {author} {\bibinfo {author} {\bibfnamefont {J.}~\bibnamefont {de~Vicente}}, \bibinfo {author} {\bibfnamefont {J.}~\bibnamefont {Stokes}},\ and\ \bibinfo {author} {\bibfnamefont {H.}~\bibnamefont {Spikes}},\ }\href {https://doi.org/10.1007/s11249-005-9067-3} {\bibfield  {journal} {\bibinfo  {journal} {Tribology Letters}\ }\textbf {\bibinfo {volume} {20}},\ \bibinfo {pages} {273} (\bibinfo {year} {2005})}\BibitemShut {NoStop}%
\bibitem [{\citenamefont {Zakhari}\ and\ \citenamefont {Bonnecaze}(2021)}]{zakhari_slip_2021}%
  \BibitemOpen
  \bibfield  {author} {\bibinfo {author} {\bibfnamefont {M.~E.~A.}\ \bibnamefont {Zakhari}}\ and\ \bibinfo {author} {\bibfnamefont {R.~T.}\ \bibnamefont {Bonnecaze}},\ }\href {https://doi.org/10.1039/D1SM00242B} {\bibfield  {journal} {\bibinfo  {journal} {Soft Matter}\ }\textbf {\bibinfo {volume} {17}},\ \bibinfo {pages} {4538} (\bibinfo {year} {2021})}\BibitemShut {NoStop}%
\bibitem [{\citenamefont {Peng}\ \emph {et~al.}(2021)\citenamefont {Peng}, \citenamefont {Serfass}, \citenamefont {Kawazoe}, \citenamefont {Shao}, \citenamefont {Gutierrez}, \citenamefont {Hill}, \citenamefont {Santos}, \citenamefont {Visell},\ and\ \citenamefont {Hsiao}}]{peng_elastohydrodynamic_2021}%
  \BibitemOpen
  \bibfield  {author} {\bibinfo {author} {\bibfnamefont {Y.}~\bibnamefont {Peng}}, \bibinfo {author} {\bibfnamefont {C.~M.}\ \bibnamefont {Serfass}}, \bibinfo {author} {\bibfnamefont {A.}~\bibnamefont {Kawazoe}}, \bibinfo {author} {\bibfnamefont {Y.}~\bibnamefont {Shao}}, \bibinfo {author} {\bibfnamefont {K.}~\bibnamefont {Gutierrez}}, \bibinfo {author} {\bibfnamefont {C.~N.}\ \bibnamefont {Hill}}, \bibinfo {author} {\bibfnamefont {V.~J.}\ \bibnamefont {Santos}}, \bibinfo {author} {\bibfnamefont {Y.}~\bibnamefont {Visell}},\ and\ \bibinfo {author} {\bibfnamefont {L.~C.}\ \bibnamefont {Hsiao}},\ }\href {https://doi.org/10.1038/s41563-021-00990-9} {\bibfield  {journal} {\bibinfo  {journal} {Nature Materials}\ }\textbf {\bibinfo {volume} {20}},\ \bibinfo {pages} {1707} (\bibinfo {year} {2021})}\BibitemShut {NoStop}%
\bibitem [{\citenamefont {Meeker}\ \emph {et~al.}(2004)\citenamefont {Meeker}, \citenamefont {Bonnecaze},\ and\ \citenamefont {Cloitre}}]{meeker_slip_2004}%
  \BibitemOpen
  \bibfield  {author} {\bibinfo {author} {\bibfnamefont {S.~P.}\ \bibnamefont {Meeker}}, \bibinfo {author} {\bibfnamefont {R.~T.}\ \bibnamefont {Bonnecaze}},\ and\ \bibinfo {author} {\bibfnamefont {M.}~\bibnamefont {Cloitre}},\ }\href {https://doi.org/10.1122/1.1795171} {\bibfield  {journal} {\bibinfo  {journal} {Journal of Rheology}\ }\textbf {\bibinfo {volume} {48}},\ \bibinfo {pages} {1295} (\bibinfo {year} {2004})}\BibitemShut {NoStop}%
\bibitem [{\citenamefont {Bongaerts}\ \emph {et~al.}(2007)\citenamefont {Bongaerts}, \citenamefont {Fourtouni},\ and\ \citenamefont {Stokes}}]{bongaerts_soft-tribology_2007}%
  \BibitemOpen
  \bibfield  {author} {\bibinfo {author} {\bibfnamefont {J.}~\bibnamefont {Bongaerts}}, \bibinfo {author} {\bibfnamefont {K.}~\bibnamefont {Fourtouni}},\ and\ \bibinfo {author} {\bibfnamefont {J.}~\bibnamefont {Stokes}},\ }\href {https://doi.org/10.1016/j.triboint.2007.01.007} {\bibfield  {journal} {\bibinfo  {journal} {Tribology International}\ }\textbf {\bibinfo {volume} {40}},\ \bibinfo {pages} {1531} (\bibinfo {year} {2007})}\BibitemShut {NoStop}%
\bibitem [{\citenamefont {Snoeijer}\ \emph {et~al.}(2013)\citenamefont {Snoeijer}, \citenamefont {Eggers},\ and\ \citenamefont {Venner}}]{snoeijer_similarity_2013}%
  \BibitemOpen
  \bibfield  {author} {\bibinfo {author} {\bibfnamefont {J.~H.}\ \bibnamefont {Snoeijer}}, \bibinfo {author} {\bibfnamefont {J.}~\bibnamefont {Eggers}},\ and\ \bibinfo {author} {\bibfnamefont {C.~H.}\ \bibnamefont {Venner}},\ }\href {https://doi.org/10.1063/1.4826981} {\bibfield  {journal} {\bibinfo  {journal} {Physics of Fluids}\ }\textbf {\bibinfo {volume} {25}},\ \bibinfo {pages} {101705} (\bibinfo {year} {2013})}\BibitemShut {NoStop}%
\bibitem [{\citenamefont {De~Vicente}\ \emph {et~al.}(2006)\citenamefont {De~Vicente}, \citenamefont {Stokes},\ and\ \citenamefont {Spikes}}]{de_vicente_soft_2006}%
  \BibitemOpen
  \bibfield  {author} {\bibinfo {author} {\bibfnamefont {J.}~\bibnamefont {De~Vicente}}, \bibinfo {author} {\bibfnamefont {J.}~\bibnamefont {Stokes}},\ and\ \bibinfo {author} {\bibfnamefont {H.}~\bibnamefont {Spikes}},\ }\href {https://doi.org/10.1016/j.foodhyd.2005.04.005} {\bibfield  {journal} {\bibinfo  {journal} {Food Hydrocolloids}\ }\textbf {\bibinfo {volume} {20}},\ \bibinfo {pages} {483} (\bibinfo {year} {2006})}\BibitemShut {NoStop}%
\bibitem [{\citenamefont {Wong}\ \emph {et~al.}(1995{\natexlab{b}})\citenamefont {Wong}, \citenamefont {Radke},\ and\ \citenamefont {Morris}}]{wong_motion_1995-1}%
  \BibitemOpen
  \bibfield  {author} {\bibinfo {author} {\bibfnamefont {H.}~\bibnamefont {Wong}}, \bibinfo {author} {\bibfnamefont {C.~J.}\ \bibnamefont {Radke}},\ and\ \bibinfo {author} {\bibfnamefont {S.}~\bibnamefont {Morris}},\ }\href {https://doi.org/10.1017/S0022112095001455} {\bibfield  {journal} {\bibinfo  {journal} {Journal of Fluid Mechanics}\ }\textbf {\bibinfo {volume} {292}},\ \bibinfo {pages} {95} (\bibinfo {year} {1995}{\natexlab{b}})}\BibitemShut {NoStop}%
\bibitem [{\citenamefont {Baudoin}\ \emph {et~al.}(2013)\citenamefont {Baudoin}, \citenamefont {Song}, \citenamefont {Manneville},\ and\ \citenamefont {Baroud}}]{baudoin_airway_2013}%
  \BibitemOpen
  \bibfield  {author} {\bibinfo {author} {\bibfnamefont {M.}~\bibnamefont {Baudoin}}, \bibinfo {author} {\bibfnamefont {Y.}~\bibnamefont {Song}}, \bibinfo {author} {\bibfnamefont {P.}~\bibnamefont {Manneville}},\ and\ \bibinfo {author} {\bibfnamefont {C.}~\bibnamefont {Baroud}},\ }\href {https://doi.org/10.1073/pnas.1211706110} {\bibfield  {journal} {\bibinfo  {journal} {PNAS}\ }\textbf {\bibinfo {volume} {110}},\ \bibinfo {pages} {859} (\bibinfo {year} {2013})}\BibitemShut {NoStop}%
\bibitem [{\citenamefont {Luo}\ \emph {et~al.}(2014)\citenamefont {Luo}, \citenamefont {Chen}, \citenamefont {Zhao}, \citenamefont {Wei}, \citenamefont {Zhao}, \citenamefont {Yue}, \citenamefont {Long}, \citenamefont {Wang},\ and\ \citenamefont {Chen}}]{luo_constriction_2014}%
  \BibitemOpen
  \bibfield  {author} {\bibinfo {author} {\bibfnamefont {Y.}~\bibnamefont {Luo}}, \bibinfo {author} {\bibfnamefont {D.}~\bibnamefont {Chen}}, \bibinfo {author} {\bibfnamefont {Y.}~\bibnamefont {Zhao}}, \bibinfo {author} {\bibfnamefont {C.}~\bibnamefont {Wei}}, \bibinfo {author} {\bibfnamefont {X.}~\bibnamefont {Zhao}}, \bibinfo {author} {\bibfnamefont {W.}~\bibnamefont {Yue}}, \bibinfo {author} {\bibfnamefont {R.}~\bibnamefont {Long}}, \bibinfo {author} {\bibfnamefont {J.}~\bibnamefont {Wang}},\ and\ \bibinfo {author} {\bibfnamefont {J.}~\bibnamefont {Chen}},\ }\href {https://doi.org/10.1016/j.snb.2014.05.028} {\bibfield  {journal} {\bibinfo  {journal} {Sensors and Actuators B: Chemical}\ }\textbf {\bibinfo {volume} {202}},\ \bibinfo {pages} {1183} (\bibinfo {year} {2014})}\BibitemShut {NoStop}%
\bibitem [{\citenamefont {Xue}\ \emph {et~al.}(2015)\citenamefont {Xue}, \citenamefont {Wang}, \citenamefont {Zhao}, \citenamefont {Chen}, \citenamefont {Yue},\ and\ \citenamefont {Chen}}]{xue_constriction_2015}%
  \BibitemOpen
  \bibfield  {author} {\bibinfo {author} {\bibfnamefont {C.}~\bibnamefont {Xue}}, \bibinfo {author} {\bibfnamefont {J.}~\bibnamefont {Wang}}, \bibinfo {author} {\bibfnamefont {Y.}~\bibnamefont {Zhao}}, \bibinfo {author} {\bibfnamefont {D.}~\bibnamefont {Chen}}, \bibinfo {author} {\bibfnamefont {W.}~\bibnamefont {Yue}},\ and\ \bibinfo {author} {\bibfnamefont {J.}~\bibnamefont {Chen}},\ }\href {https://doi.org/10.3390/mi6111457} {\bibfield  {journal} {\bibinfo  {journal} {Micromachines}\ }\textbf {\bibinfo {volume} {6}},\ \bibinfo {pages} {1794} (\bibinfo {year} {2015})}\BibitemShut {NoStop}%
\bibitem [{\citenamefont {Zilionis}\ \emph {et~al.}(2016)\citenamefont {Zilionis}, \citenamefont {Nainys}, \citenamefont {Veres}, \citenamefont {Savova}, \citenamefont {Zemmour}, \citenamefont {Klein},\ and\ \citenamefont {Mazutis}}]{zilionis_single-cell_2016}%
  \BibitemOpen
  \bibfield  {author} {\bibinfo {author} {\bibfnamefont {R.}~\bibnamefont {Zilionis}}, \bibinfo {author} {\bibfnamefont {J.}~\bibnamefont {Nainys}}, \bibinfo {author} {\bibfnamefont {A.}~\bibnamefont {Veres}}, \bibinfo {author} {\bibfnamefont {V.}~\bibnamefont {Savova}}, \bibinfo {author} {\bibfnamefont {D.}~\bibnamefont {Zemmour}}, \bibinfo {author} {\bibfnamefont {A.~M.}\ \bibnamefont {Klein}},\ and\ \bibinfo {author} {\bibfnamefont {L.}~\bibnamefont {Mazutis}},\ }\href {https://doi.org/10.1038/nprot.2016.154} {\bibfield  {journal} {\bibinfo  {journal} {Nature protocols}\ }\textbf {\bibinfo {volume} {12}},\ \bibinfo {pages} {44} (\bibinfo {year} {2016})}\BibitemShut {NoStop}%
\bibitem [{\citenamefont {Dendukuri}\ \emph {et~al.}(2005)\citenamefont {Dendukuri}, \citenamefont {Tsoi}, \citenamefont {Hatton},\ and\ \citenamefont {Doyle}}]{dendukuri_controlled_2005}%
  \BibitemOpen
  \bibfield  {author} {\bibinfo {author} {\bibfnamefont {D.}~\bibnamefont {Dendukuri}}, \bibinfo {author} {\bibfnamefont {K.}~\bibnamefont {Tsoi}}, \bibinfo {author} {\bibfnamefont {T.~A.}\ \bibnamefont {Hatton}},\ and\ \bibinfo {author} {\bibfnamefont {P.~S.}\ \bibnamefont {Doyle}},\ }\href {https://doi.org/10.1021/la047368k} {\bibfield  {journal} {\bibinfo  {journal} {Langmuir}\ }\textbf {\bibinfo {volume} {21}},\ \bibinfo {pages} {2113} (\bibinfo {year} {2005})}\BibitemShut {NoStop}%
\bibitem [{\citenamefont {Mitra}\ and\ \citenamefont {Chakraborty}(2017)}]{mitra_microfluidics_2017}%
  \BibitemOpen
  \bibfield  {author} {\bibinfo {author} {\bibfnamefont {S.~K.}\ \bibnamefont {Mitra}}\ and\ \bibinfo {author} {\bibfnamefont {S.}~\bibnamefont {Chakraborty}},\ }\href@noop {} {\emph {\bibinfo {title} {Microfluidics and {Nanofluidics} {Handbook}: {Fabrication}, {Implementation}, and {Applications}}}},\ \bibinfo {edition} {1st}\ ed.\ (\bibinfo {year} {2017})\BibitemShut {NoStop}%
\bibitem [{\citenamefont {Gupta}\ \emph {et~al.}(2016)\citenamefont {Gupta}, \citenamefont {Wang},\ and\ \citenamefont {Vanapalli}}]{gupta_microfluidic_2016}%
  \BibitemOpen
  \bibfield  {author} {\bibinfo {author} {\bibfnamefont {S.}~\bibnamefont {Gupta}}, \bibinfo {author} {\bibfnamefont {W.~S.}\ \bibnamefont {Wang}},\ and\ \bibinfo {author} {\bibfnamefont {S.~A.}\ \bibnamefont {Vanapalli}},\ }\href {https://doi.org/10.1063/1.4955123} {\bibfield  {journal} {\bibinfo  {journal} {Biomicrofluidics}\ }\textbf {\bibinfo {volume} {10}},\ \bibinfo {pages} {043402} (\bibinfo {year} {2016})}\BibitemShut {NoStop}%
\bibitem [{\citenamefont {Solomon}\ and\ \citenamefont {Vanapalli}(2014)}]{solomon_multiplexed_2014}%
  \BibitemOpen
  \bibfield  {author} {\bibinfo {author} {\bibfnamefont {D.~E.}\ \bibnamefont {Solomon}}\ and\ \bibinfo {author} {\bibfnamefont {S.~A.}\ \bibnamefont {Vanapalli}},\ }\href {https://doi.org/10.1007/s10404-013-1261-2} {\bibfield  {journal} {\bibinfo  {journal} {Microfluidics and Nanofluidics}\ }\textbf {\bibinfo {volume} {16}},\ \bibinfo {pages} {677} (\bibinfo {year} {2014})}\BibitemShut {NoStop}%
\bibitem [{\citenamefont {Marx}\ \emph {et~al.}(2016)\citenamefont {Marx}, \citenamefont {Guegan},\ and\ \citenamefont {Spikes}}]{marx_elastohydrodynamic_2016}%
  \BibitemOpen
  \bibfield  {author} {\bibinfo {author} {\bibfnamefont {N.}~\bibnamefont {Marx}}, \bibinfo {author} {\bibfnamefont {J.}~\bibnamefont {Guegan}},\ and\ \bibinfo {author} {\bibfnamefont {H.~A.}\ \bibnamefont {Spikes}},\ }\href {https://doi.org/10.1016/j.triboint.2016.03.020} {\bibfield  {journal} {\bibinfo  {journal} {Tribology International}\ }\textbf {\bibinfo {volume} {99}},\ \bibinfo {pages} {267} (\bibinfo {year} {2016})}\BibitemShut {NoStop}%
\bibitem [{\citenamefont {Nijenbanning}\ \emph {et~al.}(1994)\citenamefont {Nijenbanning}, \citenamefont {Venner},\ and\ \citenamefont {Moes}}]{nijenbanning_film_1994}%
  \BibitemOpen
  \bibfield  {author} {\bibinfo {author} {\bibfnamefont {G.}~\bibnamefont {Nijenbanning}}, \bibinfo {author} {\bibfnamefont {C.}~\bibnamefont {Venner}},\ and\ \bibinfo {author} {\bibfnamefont {H.}~\bibnamefont {Moes}},\ }\href {https://doi.org/10.1016/0043-1648(94)90150-3} {\bibfield  {journal} {\bibinfo  {journal} {Wear}\ }\textbf {\bibinfo {volume} {176}},\ \bibinfo {pages} {217} (\bibinfo {year} {1994})}\BibitemShut {NoStop}%
\bibitem [{\citenamefont {Hooke}(1988)}]{hooke_calculation_1988}%
  \BibitemOpen
  \bibfield  {author} {\bibinfo {author} {\bibfnamefont {C.~J.}\ \bibnamefont {Hooke}},\ }\href {https://doi.org/10.1115/1.3261558} {\bibfield  {journal} {\bibinfo  {journal} {Journal of Tribology}\ }\textbf {\bibinfo {volume} {110}},\ \bibinfo {pages} {167} (\bibinfo {year} {1988})}\BibitemShut {NoStop}%
\end{thebibliography}%

\clearpage
\newpage

\setcounter{figure}{0}
\setcounter{table}{0}
\makeatletter 
\renewcommand{\thefigure}{S\@arabic\c@figure}
\renewcommand{\thetable}{S\@arabic\c@table}
\makeatother

\section{Supplementary Materials}

\subsection{Microgel Production and Characterization}

Microgel beads are made by photopolymerizing droplets of a liquid pre-gel solution in a flow-focusing device~\cite{dendukuri_controlled_2005}. The continuous phase consists in fluorinated oil (FC40, 3M, France) containing 1.5 \% surfactant (Fluosurf, Emulseo, France). 
The gel precursor mixture comprises 47-56 \% v/v polyethylene-glycol (PEG, MW 200, Sigma Aldrich, Darmstadt Germany), with 9-18 \% v/v PEG-diacrylate (PEGDA, MW 700, Sigma Aldrich, Darmstadt Germany), such that 65 \% of volume is a combination of PEG and PEGDA, 5 \% v/v photoinitiator (2-hydroxy-2-methyl-1-phenyl-propan-1-one, Sigma Aldrich, Darmstadt Germany), as well as 1 \% w/v sodium dodecyl sulfate surfactant (SDS, Sigma Aldrich, Darmstadt Germany) in an aqueous solution. The flow rates of the oil and aqueous phases are adjusted to obtain beads with diameters in the range $d \in (80 , 140) \; \mu \rm m$, summarized in Tab.~\ref{tab:sup_Resistance_GelChar}. The droplets  gel as they pass under a UV illumination while flowing in the microchannel. Microgels are collected out of the microfluidic device and  washed several times in ethanol and water before being put in a bath of water with 1 \% SDS to prevent them from sticking together. In the case of experiments using glycerol as a  carrier fluid, the microgels are left in a bath of glycerol. Finally, the microgels are allowed to soak at least overnight.

\begin{table}[!htb]
    \centering
    \begin{tabular}{|c|c|c|}
    \hline
     PEGDA \% aq (v/v)    &  $d$ ($\mu m$) & $E^{\star}$ ($kPa$)\\
     \hline
      $9$   & $85.3 \pm 2.5$ & $17.8\pm5.1$\\
      $9$ & $104.9 \pm 2.2$ & $6.52\pm1.20$\\
      $9$ & $124.6\pm 4.8$ & $6.17\pm3.33$\\
      $12$ & $97.3 \pm 2.6$ & $16.1\pm1.7$\\
      $12$ & $108.1 \pm 3.0$ & $19.0\pm12.8$\\
      $18$ & $109.5 \pm 4.4$ & $154\pm125$\\
      \hline
    \end{tabular}
    \caption{
    Summary of PEG microgel diameters and elastic moduli used. Diameters, $d$, were measured optically, and effective elasticity, $E^{\star}$ were measured using micro-constriction aspiration. Uncertainty represents 1 standard deviation of a sample of $n = 3$.
    }
    \label{tab:sup_Resistance_GelChar}
\end{table}

The effective elasticity, $E^{\star}$, of the microgels is measured using a method developed by Moore \emph{et al.} we've dubbed micro-constriction aspiration~\cite{moore_clogging_2023}.
In short, a microgel bead is flowed into a microfluidic constriction of width $w$, as shown in Fig.~\ref{fig:sup_MicroConstAsp}a. Once in the constriction, the bead partially plugs the thrupass channel, forcing fluid to pass through a bypass line and away from the thrupass line, with known hydrodynamic resistance, $R_{\rm byp}$ and $R_{\rm thru}$ respectively. The carrier fluid is seeded with tracer particles, and the proportion of fluid passing through the bypass is measured by particle tracking velocimetry. We can then calculate the fluid flow rates through the bypass and constriction, $Q_{\rm byp}$ and $Q_{\rm thru}$ respectively. The pressure $P_{\rm const}$ applied to the trapped bead is calculated as
\begin{equation}
    P_{\rm const} = Q_{\rm thru} \left(\frac{Q_{\rm thru}}{Q_{\rm byp}}R_{\rm byp} - R_{\rm thru} \right).
    \label{eq:sup_Pconst}
\end{equation}

The microfluidic constriction is chosen such that the microchannel height matches the gel diameter, $h\approx d$. This allows for the 3-dimensional geometry of the bead to be known while undergoing deformation. The elongation $\delta$ of the bead in the constriction, and the angle $\theta$ the bead makes with the corners of the constriction, are measured as shown in Fig.~\ref{fig:sup_MicroConstAsp}b. Additionally, the radius of the constriction corners, $r_{\rm c}$, are also measured during experiments. The elasticity of the bead is estimated by solving for the radius of contact
\begin{equation}
    \label{eq:sup_Egel}
    \begin{aligned}
    a^3   = 
    \\
    \frac{3P_{\rm const}d}{4\pi E^{\star}\sin{\theta}}\left[aw+\left(\frac{d}{3}-\frac{2a}{3}\right)\left(w+2r_c(1-\cos{\theta})\right)\right]
    .
    \end{aligned}
\end{equation}
The solution in Eq.~\eqref{eq:sup_Egel} is then equated with the solution for the contact radius based on the bead displacement
\begin{equation}\label{eqn:sup_ContactBackfull}
    \delta = \frac{2a^2}{d\sin{\theta}}\mathcal{G}(\mathcal{Z}),
\end{equation}
where $\mathcal{G}=\mathcal{Z}^{-2}\left(1-\frac{\mathcal{E}(\mathcal{Z})}{\mathcal{K}(\mathcal_Z)}\right)^{-1}$, $\mathcal{Z}$ is the eccentricity of contact, and $\mathcal{K}(\mathcal{Z})$ and $\mathcal{E}(\mathcal{Z})$ are the complete elliptical integral of the first and second kind respectively. The eccentricity of contact is estimated by $\frac{2r_c}{2r_c+d}\approx \left(1-\mathcal{Z}^2\right)\left(\ln{\frac{\sqrt{1-\mathcal{Z}^2}}{4}}-1\right)$ ~\cite{moore_clogging_2023}.
Multiple pressures are tested for a single bead, and the best fit solution to Eq. ~\eqref{eq:sup_Egel} is used as the value of the gel elasticity $E^{\star}$. A minimum of $n=6$ beads are used in estimating the overall elasticity of each bead batch. The measured elastic moduli are presented in Tab.~\ref{tab:sup_Resistance_GelChar}.

\begin{figure}
    \centering
    \includegraphics{Figures/Figure_SupFig_MicrogapAsp_V03.pdf}
    \caption{
    (a) Top: diagram showing the microfluidic channel used to trap and deform a microgel bead, enabling to measure its elasticity $E^{\star}$. Bottom: micrograph of a gel bead deforming into the constriction channel of width $w$.
    (b) A diagram of a deformed bead as seen from above. 
    Adapted from Moore \emph{et al.} \cite{moore_clogging_2023}.
    }
    \label{fig:sup_MicroConstAsp}
\end{figure}

\subsection{Microfluidic Comparator}

\begin{figure}
    \centering
    \includegraphics{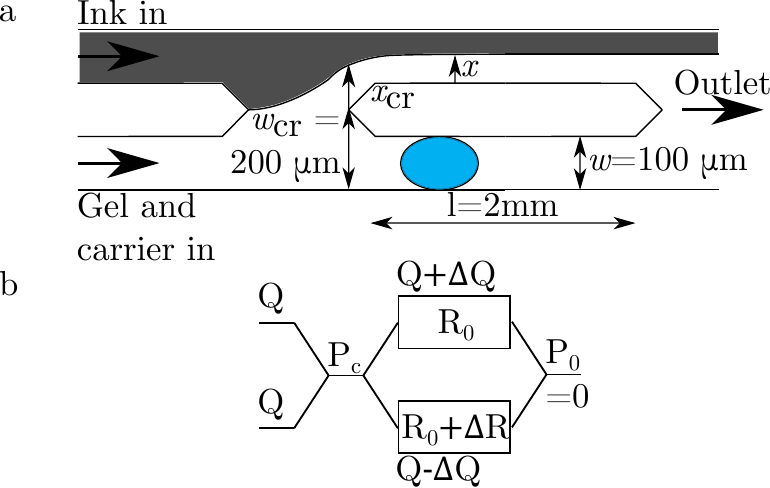}
    \caption{
    (a) A diagram showing the microfluidic comparator in operation. The deflections of the ink border are highlighted at $x$ and $x_{\rm cr}$.
    (b) A resistance diagram representing the relationship between pressure and flow in the microfluidic comparator when a bead is present.
    }
    \label{fig:sup_comparator}
\end{figure}

To measure the resistance of a bead passing through a constriction, particles are sent through a microfluidic comparator~\cite{vanapalli_hydrodynamic_2009}, 
 see Fig.~\ref{fig:sup_comparator}a. The channel is  made of polydimethylsyloxane (PDMS, SYLGARD 184, Sigma Aldrich, Darmstadt, Germany) using standard soft lithography techniques~\cite{mitra_microfluidics_2017}. Syringe pumps (Nemesis, Cetoni GmbH, Korbussen, Germany) are used to inject a dilute suspension of beads at one inlet, and a dyed fluid at the other. The two inlets join, then separate again into two distinct channels, called the reference channel (at the top, with the dyed fluid) and the test channel (at the bottom, with the gel bead), see Fig.~\ref{fig:sup_comparator}a. As the bead passes into the test channel the resistance in this channel increases, forcing fluid into the reference line, deflecting the ink boundary a distance of $x$ in the channel, and $x_{\rm cr}$ at the channel entrance, see Fig.~\ref{fig:sup_comparator}a. The deflection $x_{\rm cr}$ is directly related to the added resistance due to the gel bead. The test channel measures a width and height of $w=100\; \mu \rm m$ and $h=95\; \mu \rm m$ respectively, with a length of $l=2\; \rm mm$. The entry to the channel, at the junction, measures $w_{\rm cr}=200\; \mu \rm m$.

The added resistance to flow imposed by the moving bead is determined from the resistance diagram shown in Fig.~\ref{fig:sup_comparator}b. At the junction, the pressure drop is $P_c-P_0=P_c$, where we use $P_0$ as our origin of pressures. The resistance in the test channel, $R_0+\Delta R$, can then be determined from the inlet flow rate, $Q$, and the flow deflected to the bypass line, $\Delta Q$, by:
\begin{equation}
    P_c= \left( Q - \Delta Q \right)\left( R_0 + \Delta R \right)=\left( Q + \Delta Q \right)R_0. 
    \label{eq:sup_Rbalance1}
\end{equation}
Solving for the added resistance due to the bead, $\Delta R$, we find:
\begin{equation}
    \Delta R = R_0 \left( \frac{Q+\Delta Q}{Q-\Delta Q}-1 \right).
    \label{eq:sup_Rbalance2}
\end{equation}
We assume the pressure drop $P_{\rm gel}$ due to the presence of a confined bead is local to the bead itself, such that
\begin{equation}
    P_{\rm gel}= \Delta R \left( Q - \Delta Q \right).
    \label{eq:sup_Rbalance3}
\end{equation}
Combining Eqs.~\eqref{eq:sup_Rbalance2} and ~\eqref{eq:sup_Rbalance3}, we see that the pressure acting on the bead can be expressed as
\begin{equation}
    P_{\rm gel}=2 R_{\rm 0} \Delta Q.
\end{equation}

The hydrodynamic resistance $R_0$ of a rectangular channel is calculated from the Poiseuille equation as $R_0=28.5\frac{\mu l}{w^2h^2}$ for a fluid of dynamic viscosity $\mu$.
The deflected flow rate is calculated using the approximation presented in Vanapalli \emph{et al.}~\cite{vanapalli_hydrodynamic_2009}:
\begin{equation}
    \frac{\Delta Q}{Q}=\frac{\int_{-w/2}^{-w/2+x}\int_{-h/2}^{h/2}u(x,y)dydx}{\int_{-w/2}^{w/2}\int_{-h/2}^{h/2}u(x,y)dydx},
\end{equation}
where the local velocity is approximated by $u(x,y) \approx \left(1-\left(\frac{2y}{h}\right)^2\right)\left(1-\left(\frac{2x}{w}\right)^m\right)$, and $m=\sqrt{2}(h/w)+0.89(w/h)$. 
Empirical testing reveals a proportionality of $x/w=x_{\rm cr}/w_{\rm cr}$, and as such $\Delta Q$ is estimated using $x_{\rm cr}$ for precision, which is consistent with previous work using microfluidic comparators ~\cite{vanapalli_hydrodynamic_2009}. 

In addition to measuring the added resistance from a bead, the microfluidic comparator is also used to determine the viscosity of the fluids used here \cite{gupta_microfluidic_2016,solomon_multiplexed_2014}. Noting that in a rectangular microchannel, $R_0\propto \mu$, we adjust the inlet flow rates such that the dyed ink line lands at exactly $x_{\rm cr}=0$. 
The inlet flow rates are then related to the fluid viscosities through 
\begin{equation}
    \frac{Q_2}{Q_1}=\frac{\mu_1}{\mu_2} f \left(\frac{2w_{cr}}{h}, \frac{\mu_2}{\mu_1}\right) ,
    \label{eq:SolomonVanapaliComparator}
\end{equation}
where the subscripts $1$ and $2$ represent the two fluids. $f$ is a function of the viscosity ratios and channel aspect ratio at the comparator described in Solomon and Vanapalli \cite{solomon_multiplexed_2014}, where
\begin{equation}
    f = \frac{\frac{w_{\rm cr}}{48h}-\sum_n^\infty{\frac{\left(1-(-1)^n\right) \left( \tanh{\frac{n \pi w_{\rm cr}}{h}} + \xi \left( 1-1/\cosh{\frac{n \pi w_{\rm cr}}{h}} \right) \right)}{n^5 \pi^5}}}{\frac{w_{\rm cr}}{48h}+\sum_n^\infty{\frac{\left(1-(-1)^n\right) \left( - \tanh{\frac{n \pi w_{\rm cr}}{h}} + \xi \left( 1-1/\cosh{\frac{n \pi w_{\rm cr}}{h}} \right) \right)}{n^5 \pi^5}}}
\end{equation}
and where $\xi=\frac{\left( \frac{\mu_2}{\mu_1} - 1 \right)\left( 1-\frac{1}{\cosh{n \pi w_{\rm cr}/h}} \right)}{\left(-1-\frac{\mu_2}{\mu_1} \tanh{n\pi w_{\rm cr}/h} \right)}$.
Using pure water at room temperature as a reference fluid ($\mu_{\rm H2O} = 1.00*10^{-3} \; \rm Pa.s$), the viscosity of the water-glycerol solution used here is measured to be $\mu_{\rm glyc}=2.24*10^{-2} \;\rm Pa.s$ This methodology is also used to verify that the addition of dye to the fluid has negligible effect on the viscosity of the carrier fluids.

\subsection{bead Velocity}

\begin{figure}
    \centering
    \includegraphics{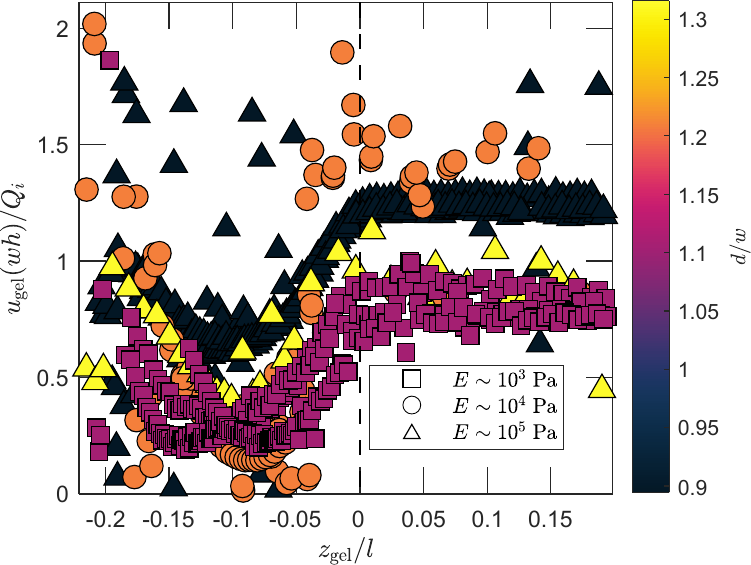}
    \caption{
    The velocity, $u_{\rm gel}$, of a sample of 13 beads as they pass through the length of the comparator, normalized to the fluid velocity, $Q_i/wh$, compared with the position of the bead as they enter the comparator, $z_{\rm gel}/l$. The position $z_{\rm gel}=0$ corresponds to the point where the comparator width becomes constant. Relative size of the beads is shown in the color-bar. All samples taken from tests using water as the carrier fluid.
    }
    \label{fig:sup_Velocitylinear}
\end{figure}

The analysis of lubrication between the bead and microchannel walls assumes a steady state, where the pressure and friction forces are in equilibrium, \emph{i.e.} $F_{\rm P} = F_{\rm f}$. 
This can be verified by noting that beads passing through the constriction of the comparator move at a constant velocity, as seen in the sample bead trajectories in Fig.~\ref{fig:sup_Velocitylinear}. 
While beads approaching the entry of the comparatory move with relatively unsteady velocities, once they enter into the comparator at $z_{\rm gel}=0$, the beads have accelerated to within a 95 \% confidence interval of their mean value.
This is the case in all samples studied, regardless of fluid, bead elasticity, flow rate or diameter. It should also be noted that while the total length of the comparator is $l=2 \rm \; mm$, the high speed videos taken of bead passage comprised only approximately the 1st quarter of this length. Due to the uniformity of the microchannel, there is no reason to believe the bead velocity changes when it leaves the field of view of the microscope camera.

\subsection{Gutter Friction}

\begin{figure}
    \centering
    \includegraphics{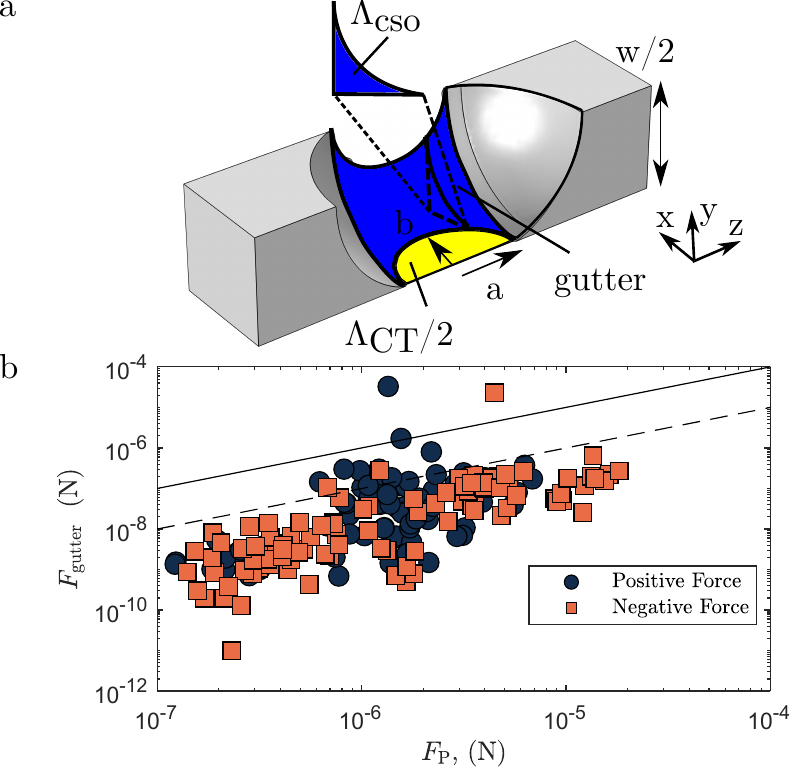}
    \caption{
    (a) A diagram showing the gutter around the bead, highlighted in blue, around the bead body (shown as a quarter section along the axes of symmetry). The cross section of gutter, where fluid passes the gel, is highlighted as $\Lambda_{\rm CSO}$, as is the radius of the gel-wall contact, $a$.
    (b) The estimated pressure force, $F_{\rm P}$, as well as the estimated gutter friction, $F_{\rm gutter}$. Force acting on the bead in the direction of flow are marked by open circles, while gutter friction against the direction of flow is represented by x. The lines $y=x$ and $y=\frac{1}{10}x$ are provided for reference.
    }
    \label{fig:sup_GutterFriction}
\end{figure}

Friction acting on a flowing bead can be separated into two sections, the lubricating region, where the bead is nearly in contact with the microchannel walls, and the gutters, accounting for all the rest of the surface area around the bead. The main analysis in the text focuses on the friction occurring in the lubricating region. Here we also account for friction occurring in the gutters between the bead and microchannel.

To analyze flow through the gutters, we must first define clearly what is meant by this geometry. The area of the gutters, $\Lambda_{\rm CSO}$ is shown in Fig.~\ref{fig:sup_GutterFriction}a. 
At its smallest point, this area is the area left between the square microchannel of section $w^2$ and the gel of cross-section $\Lambda_{\rm CS}$. The cross-sectional area of a gutter is then $\Lambda_{\rm CSO,0}=\frac{1}{4}\left(w^2-\Lambda_{\rm CS}\right)$. 

We now make several approximations to simplify the analysis. In the absence of flow, the area of contact between the compressed gel and the channel wall is $\lct$, and has an ellipsoidal shape of major axis $a$ and minor axis $b$. Simulation results of $\lct$ show that $|\frac{a-b}{a}|<0.25$, and therefore we approximate $\Lambda_{\rm CT} \approx \pi a^2$. We assume that the length of the gutter is $a$. We do not consider changes in the cross-sectional area of the gutter, and scale our analysis with the minimal value of the cross-sectional area, taking $\Lambda_{\rm CSO}=\Lambda_{\rm CSO,0}$. Last, we approximate the profile of the bead in this gutter area as a quarter circle, and so have $\Lambda_{\rm CSO,0}=\left(w-b\right)^2 \left( 1 - \frac{\pi}{4} \right)$.

Next, we consider the flow passing by the bead. In the frame of reference of the gel, the channel walls are moving backwards at $u_{\rm wall}'=u_{\rm gel}$. From this standpoint, by conservation of volume, the flow rate past the bead is therefore $Q_{\rm gutter}=(u_{\rm fluid}-u_{\rm gel})w^2$. 
The mean velocity in the gutters is $u_{\rm gutter}'=Q_{\rm gutter}/\Lambda_{\rm CSO}$.

The shear stress acting on the surface of the bead is defined as 
\begin{equation}
    \sigma_{\rm gel}=\mu \left.\frac{\partial u}{\partial z}\right|_{\Lambda_{\rm gel}},    
\end{equation}
where $\sigma_{\rm gel}$ and $u$ are perpendicular to the surface of the bead, and $z$ is perpendicular, allowing $\Lambda_{\rm gel}$ to be the surface of the bead in the gutter.

The flow in the gutter is due to the pressure difference between the fore and aft of the gel, and to the gel bead entraining fluid with it, leading to  a Couette-Poiseuille flow. The full solution of the Couette-Poiseuille flow profile in this geometry is  complicated, and we use a scaling law analysis to estimate the shear stress on the bead:
\begin{equation}
    \sigma_{\rm gel} \sim \frac{\mu u_{\rm gutter}}{\sqrt{\Lambda_{\rm CSO}}}.
    \label{eq:sup_sigmagel_sim}
\end{equation}

To estimate the shear force on the bead, we take the approximation mentioned above that $\Lambda_{\rm CSO}=\Lambda_{\rm CSO,0}$, for the entire length of the gutter. Taking the curve of the bead surface to be circular, the total surface area of the gutter (for all 4 gutters combined) is $\Lambda_{\rm surf}\approx 2 \pi a(w-2a)$, noting from simulation results that $\lct \approx \pi a^2$ for beads tested. With that in mind, the friction force in the gutters is approximated by
\begin{equation}
    F_{\rm gutter} \sim \pi a(w-2a) \frac{\mu u_{\rm gutter}}{\sqrt{\Lambda_{\rm CSO}}}.
    \label{eq:sup_Fgutter_sim}
\end{equation}

The gutter friction $F_{\rm gutter}$ can be seen compared with the pressure force acting on the bead, $F_{\rm P}=P_{\rm gel}\lct$, in Fig.~\ref{fig:sup_GutterFriction}b. With only a handful of outliers, the magnitude of the gutter friction is less than a tenth that of the pressure force. The gutter friction is also bidirectional, pushing the bead faster or slowing it down depending on the specific combination of parameters, however most of the beads are experiencing a weak resistance to flow due to gutter friction. This is simply an effect of whether the mean relative gutter flow is positive or negative. It should also be noted that the estimate here likely considerably overestimates the magnitude of the friction in the gutter due to geometric considerations. By assuming that the bead side of the gutter is shaped like a quarter cylinder, the shear is kept at its maximum value for the length of the gutter. While this analysis does neglect the end caps of the bead on either side of the gutter, this friction can be expected to be considerably less than that within the gutter, as $|u_{\rm gutter}|$ decreases and therefore so does the shear.

\subsection{Scaling Analysis of Friction}

As described in the main text, as the bead passes through the microchannel, a thin film of fluid develops between the two surfaces where the bead would otherwise be in contact with the microchannel. This lubricating film, acting over the area $\lct$, is due to the lubrication pressure $P(x,z)$ pushing the bead away from the wall surface a distance of $t(x,z)$. The coupling of the lubrication pressure with the gel elasticity comes to describe the phenomena of elastohydrodynamics~\cite{zargari2007investigation, hamrock_isothermal_1976, hamrock_fundamentals_2004}. In the case of soft-elastohydrodynamics, the thickness of the lubrication layer $t(x,z)$ is described by the Hertz equation, such that:
\begin{equation}
    t(x,z)=t_{\rm 0}+2 \frac{\pi}{E^{\star}} \int_{\lct}{\frac{P(x',z')}{\sqrt{(x-x')^2+(z-z')^2}}}dx'dz'.
    \label{eq:sup_Hertz}
\end{equation}
In this case, $t_0$ represents the undeformed shape of the bead. Because the bead is considerably softer than the PDMS walls, the elasticity $E^{\star}$ is that of the bead. Meanwhile, the pressure in the lubricating film, $P(x,z)$ is described by the Reynolds equation~\cite{guyon2015physical}:
\begin{equation}
    \frac{\partial}{\partial x}\left(t^3\frac{\partial P}{\partial x} \right)+\frac{\partial}{\partial z}\left( t^3 \frac{\partial P}{ \partial z} \right)=6\mu u_{\rm gel}\frac{\partial t}{\partial x}.
    \label{eq:sup_Reynolds}
\end{equation}

The coupled Eqs.~\eqref{eq:sup_Hertz} and \eqref{eq:sup_Reynolds} must be solved numerically, and the precise solution has been shown to depend upon thermal, fluid and solid properties of the elastohydrodyanmic system \cite{hamrock_isothermal_1976,marx_elastohydrodynamic_2016,nijenbanning_film_1994,hooke_calculation_1988,sadowski_friction_2019,de_vicente_frictional_2005,cartas_ayala_hydrodynamic_2013,khan_passage_2017}. This analysis can be simplified by considering the scaling of the pressure and thickness in the lubricating film. This simplified scaling analysis is described below.

In order to understand the typical scaling of the lubricating thickness, we define the typical pressure within the lubrication domain, $\bar{P}$, and consider how this affects the typical thickness of the lubricating layer, $\bar{t}$. When considering the Hertzian deformation of the bead, we must first note that the area of the lubricating domain is approximately that of dry contact between the bead and microchannel, such that $\lct \sim a^2$, where $a$ is the radius of the lubricating film. With this in mind, the scaling of Eq.~\eqref{eq:sup_Hertz} is:
\begin{equation}
    \bar{t} \sim \frac{\bar{P}a^2}{E^{\star} a},
    \label{eq:sup_HertzScale}
\end{equation}
which provides the scaling for the typical pressure in the lubricating film:
\begin{equation}
    \bar{P}\sim E^{\star}\frac{\bar{t}}{a}.
    \label{eq:sup_scalingPressure}
\end{equation}

Likewise, Eq.~\eqref{eq:sup_Reynolds} writes as a scaling law:
\begin{equation}
    \frac{\bar{t}^3 \bar{P}}{a^2} \sim \mu u_{\rm gel} \frac{\bar{t} }{a},
    \label{eq:sup_ReynoldsScale}
\end{equation}
which, combined with Eq.~\eqref{eq:sup_scalingPressure} leads to the following scaling for the lubricating thickness:
\begin{equation}
    \bar{t}\sim\left(\frac{\mu u_{\rm gel}}{E^{\star}}\right)^{1/3}a^{2/3}.
    \label{eq:sup_scalingThickness}
\end{equation}
The typical order of magnitude for $\bar{t}$ is therefore $\bar{t} \sim 1 \;  \mu \rm m$, using the representative values $E^{\star}=10 \;\rm kPa$, $\mu = 1 \;\rm mPa.s$, $u_{\rm gel}=0.1 \; \rm m/s$, and $a=25 \;   \mu \rm m$.

The thin lubricating layer causes a shear stress, $\sigma$, to act on the surface of the bead. This shear stress results in a friction acting over the lubricating surface, $\lct$, such that
\begin{equation}
    F_{\rm f} = \int_{\Lambda_{\rm CT}} \sigma(x,z) dxdz.
    \label{eq:sup_ReynoldFric}
\end{equation}

The shear stress scales as 
\begin{equation}
    \sigma  \sim \frac{\mu u_{\rm gel}}{\bar{t}}, 
    \label{eq:sup_ReynoldShear}
\end{equation}
and therefore the friction on the bead surface scales as
\begin{equation}
    F_f\sim \frac{\mu u_{\rm gel} \lct }{\bar{t}} .
    \label{eq:FricScale}
\end{equation}
Finally, the scaling for the lubricating thickness described in Eq.~\eqref{eq:sup_scalingThickness} can be expanded, noting that $\lct = \pi a^2$, such that the lubricating friction is expected to scale as 
\begin{equation}
    F_f\sim (\mu u_{\rm gel} \lct)^{2/3}{E^{\star}}^{1/3}.
    \label{eq:FricScale2}
\end{equation}

\end{document}